# Ecosystem-bedrock interaction changes nutrient compartmentalization during early oxidative weathering

Dragos G. Zaharescu[1,2,3], Carmen I. Burghelea[3], Katerina Dontsova[3,4], Jennifer K. Presler[3], Edward A. Hunt[3], Kenneth J. Domanik[5], Mary K. Amistadi[6], Shana Sandhaus[3,7], Elise N. Munoz[3,7], Emily E. Gaddis[3,8], Miranda Galey[3,9], Maria O. Vaquera-Ibarra[3,10], Maria A. Palacios-Menendez[3,11], Ricardo Castrejón-Martinez[3,12], Estefanía C. Roldán-Nicolau[3,12], Kexin Li[3,13], Raina M. Maier[4], Christopher T. Reinhard[1,2], Jon Chorover[3,4]

[1]Department of Earth and Atmospheric Sciences, Georgia Institute of Technology, Atlanta, GA, U.S.A. [2]Alternative Earths Team, NASA Astrobiology Institute, University of California, Riverside, CA, U.S.A., [3]Biosphere 2, The University of Arizona, Tucson, AZ, U.S.A. [4]Department of Environmental Science, The University of Arizona, Tucson, AZ, U.S.A. [5]Lunar and Planetary Laboratory, The University of Arizona, Tucson, AZ, U.S.A. [6]Arizona Laboratory for Emerging Contaminants, The University of Arizona, Tucson, AZ, U.S.A. [7]Honor's College, The University of Arizona, Tucson, AZ, U.S.A. [8]Williams College, Williamstown, MA, U.S.A. [9]Biology Department, The University of Minnesota, Duluth, MN, U.S.A. [10]University of the Americas Puebla, Puebla, México. [11]The University of Caribe, Cancún, México. [12]Universidad Nacional Autonoma de Mexico. [13]Department of Computer Sciences, University of Wisconsin-Madison, WI, U.S.A.

Correspondence and requests for materials should be addressed to D.G.Z. (email: zaha_dragos@yahoo.com).

**Ecosystem-bedrock interactions power the biogeochemical cycles of Earth's shallow crust, supporting life, stimulating substrate transformation, and spurring evolutionary innovation. While oxidative processes have dominated half of terrestrial history, the relative contribution of the biosphere and its chemical fingerprints on Earth's developing regolith are still poorly constrained. Here, we report results from a two-year incipient weathering experiment. We found that the mass release and compartmentalization of major elements during weathering of granite, rhyolite, schist and basalt was rock-specific and regulated by ecosystem components. A tight interplay between physiological needs of different biota, mineral dissolution rates, and substrate nutrient availability resulted in intricate elemental distribution patterns. Biota accelerated $CO_2$ mineralization over abiotic controls as ecosystem complexity increased, and significantly modified stoichiometry of mobilized elements. Microbial and fungal components inhibited element leaching (23.4% and 7%), while plants increased leaching and biomass retention by 63.4%. All biota left comparable biosignatures in the dissolved weathering products. Nevertheless, the magnitude and allocation of weathered fractions under abiotic and biotic treatments provide quantitative evidence for the role of major biosphere components in the evolution of upper continental crust, presenting critical information for large-scale biogeochemical models and for the search for stable *in situ* biosignatures beyond Earth.**

Vast nutrient and energy transfers between Earth's geological, hydrological and atmospheric reservoirs support the evolution of terrestrial life and surface habitability. On modern Earth, this massive, but finely-tuned bioreaction continuously consumes reactive minerals, oxygen and atmospheric acidity (as rainwater-dissolved $CO_2$) to drive the cycle of most chemical elements through hydrolytic and oxidative weathering. At the catchment scale, weathering of rock to soil and sediment prepares the exposed surface for developing ecosystems by physically breaking and chemically transforming rock into regolith, liberating macro and micronutrients from primary minerals and transporting rock-derived solutes into flowing water, incorporating these products into novel mineral-organic aggregates, and making them accessible to uptake by biota. The structure of colonizing ecosystems depend on the intensity and trajectory of these processes, on the composition of nutrient pools in the bedrock, and their feedback relationships with the wider atmosphere and climate[1,2].

From Vernadsky's early recognition of the crucial role of Earth's biosphere in crustal transformation[3], to recent efforts to resolve Critical Zone function[4], ecotope development[5], and biotic impacts on mineral diversification through time[6], there has been a gradually improving understanding of life's role in the biogeochemical evolution of Earth's surface. However, whether abiotic processes dominate over biotic or to what extent different components of the living world contribute and have contributed to the evolution of Earth's upper continental crust remains poorly constrained.

An interplay between thermal/oxidation/hydraulic fracturing and abiotic dissolution reactions (oxidation and hydrolysis) initiates mineral weathering. In the absence of plants, autotrophic microbes — the first mineral colonizers and an abundant component of the terrestrial biosphere — start biological weathering by oxidizing redox-sensitive elements (Fe and Mn) and fixing $CO_2$ into biomass. They also release $CO_2$ during aerobic respiration, along with secondary metabolites, such as organic acids, surfactants and siderophores, that enhance the extraction and sequestration of life-limiting nutrients such as Fe and P[7–9] and provide a reduced C source for eventual substrate colonization by heterotrophic organisms. While these interactions can accelerate weathering, they can also inhibit it by coating reactive surfaces with biopolymers[10].

Plant growth adds to microbial effects through biomechanical (e.g., fracturing) and biochemical weathering[11–13]. Plants fix $CO_2$ from the atmosphere via photosynthesis, transferring it to the rhizosphere as reduced C compounds including low molecular mass organic acids that, together with $CO_2$ produced during microbial and root respiration, increase the proton and ligand donor pool and accelerate weathering reactions[14]. Roots also release border cells (an immunity mechanism)[15] and other cortex exfoliates, which complex mineral-derived ions, nurture microbial communities, and induce changes in both the speciation and overall abundance of elements dissolved in pore fluids[16]. The capacity of most land plants to assimilate $CO_2$ and nutrients has evolved in strong interdependence with mycorrhizae, a co-evolved plant-fungal symbiosis, which historically enabled vascular plants to colonize the land surface[17].

Complex interactions at the biotic-abiotic interface, therefore, have the potential to change the stoichiometry of weathering and are thus critical drivers of Earth's global carbon, nitrogen and lithogenic element cycles, controlling both the extent of soil and ecosystem habitability on the planet and the capacity for regolith to develop and store *in situ* biosignatures. The fact that ecosystem health is directly tied to bedrock composition is readily illustrated, e.g., through detrimental responses associated with external perturbations such as N input[18].

Elements released from rock weathering are subject to several competing fates: removal in flowing water, complexation with organic matter or mineral surfaces, incorporation into secondary minerals, or uptake into growing biota. During incipient weathering, such element mass distributions are controlled by their availability in the parent rock, the rates of mineral dissolution, the needs/selectivity of the biota, and the relative saturation of solutions with respect to secondary mineral phases. In actively uplifting and eroding topography (e.g. mountains), a larger supply of fresh protolith in contact with rain water and higher erosion compared to quiescent landscapes trigger comparatively greater aqueous denudation[19] and exponentially increased soil production rates[20]. Fresh protolith, typically in large disequilibrium with surface Earth conditions, is therefore among the most bio-reactive terrestrial substrates and is critical in initiating local and global nutrient cycles.

While mass balance approaches can, in principle, trace element pathways and material fluxes in an evolving biosphere, the relative contributions of Earth's abiotic and biotic components, including microorganisms, plants, and their associated mycorrhizae, are challenging to disentangle in field systems. To begin closing this gap, we performed a two-year controlled mesocosm experiment (Fig. s1). We tested the hypothesis that major element compartmentalization among neo-formed solids, aqueous solutions, and biomass during incipient weathering of basalt, granite, rhyolite, and schist depends significantly and differentially on the evolutionary emergence and activity of microbes (B, bacteria treatment), vascular plants (BG, bacteria-grass treatment), and arbuscular mycorrhiza (BGM, bacteria-grass-arbuscular mycorrhiza treatment) relative to the abiotic dissolution of the same substrates (C- control treatment). Based on differential nutrient requirements of plants and microbes we also hypothesized that less P and Mg would be lost from the system compared to other ions, and this would reflect as potential biosignatures. Aqueous geochemical denudation, element allocation to plant biomass, and changes in labile solid-phases were measured in biotic and abiotic treatments over the course of two years. The use of a stepwise treatment design and unreacted geomedia of consistent particle size offered the advantage of greatly simplifying the inherent complexity of an evolved natural weathering environment, allowing us to better constrain the impacts of different biotic treatments on mass balance and the production and retention of *in situ* biosignatures.

## Results and discussion

**Substrate characterization.** Analysis of the four fresh rocks, ground and isolated to give consistent particle size distributions prior to the experiment (see methods) revealed distinct physical, mineralogical and chemical properties (Fig. 1, Fig. s2, Ref.[21,22]). Bulk column density, which is important for plant rooting, decreased with increasing pore volume (p.v.) according to: granite ($1.34 \pm 0.003$ g*cm$^{-3}$, p.v. = $32 \pm 0.8$ %) > basalt ($1.32 \pm 0.03$g*cm$^{-3}$, p.v. = $33 \pm 0.7$ %) > rhyolite ($1.24 \pm 0.02$ g*cm$^{-3}$, p.v. = $35 \pm 1$ %) > schist ($1.04 \pm 0.02$ g*cm$^{-3}$, p.v. = $43 \pm 1$ %).

Basalt was rich in amorphous volcanic glass with crystalline inclusions of Ca-Mg-Al pyroxenes, Mg-olivine, and Ca-feldspar (Fig. 1; Ref.[22]). Rhyolite was dominated by Na/K feldspars and quartz. Similarly, the granite substrate mainly contained quartz, Na and K feldspar, and smaller amounts of Mg- and Fe-rich biotite. Schist comprised quartz and K, Al, Mg, and the Fe-rich micas muscovite, phengite and biotite, which

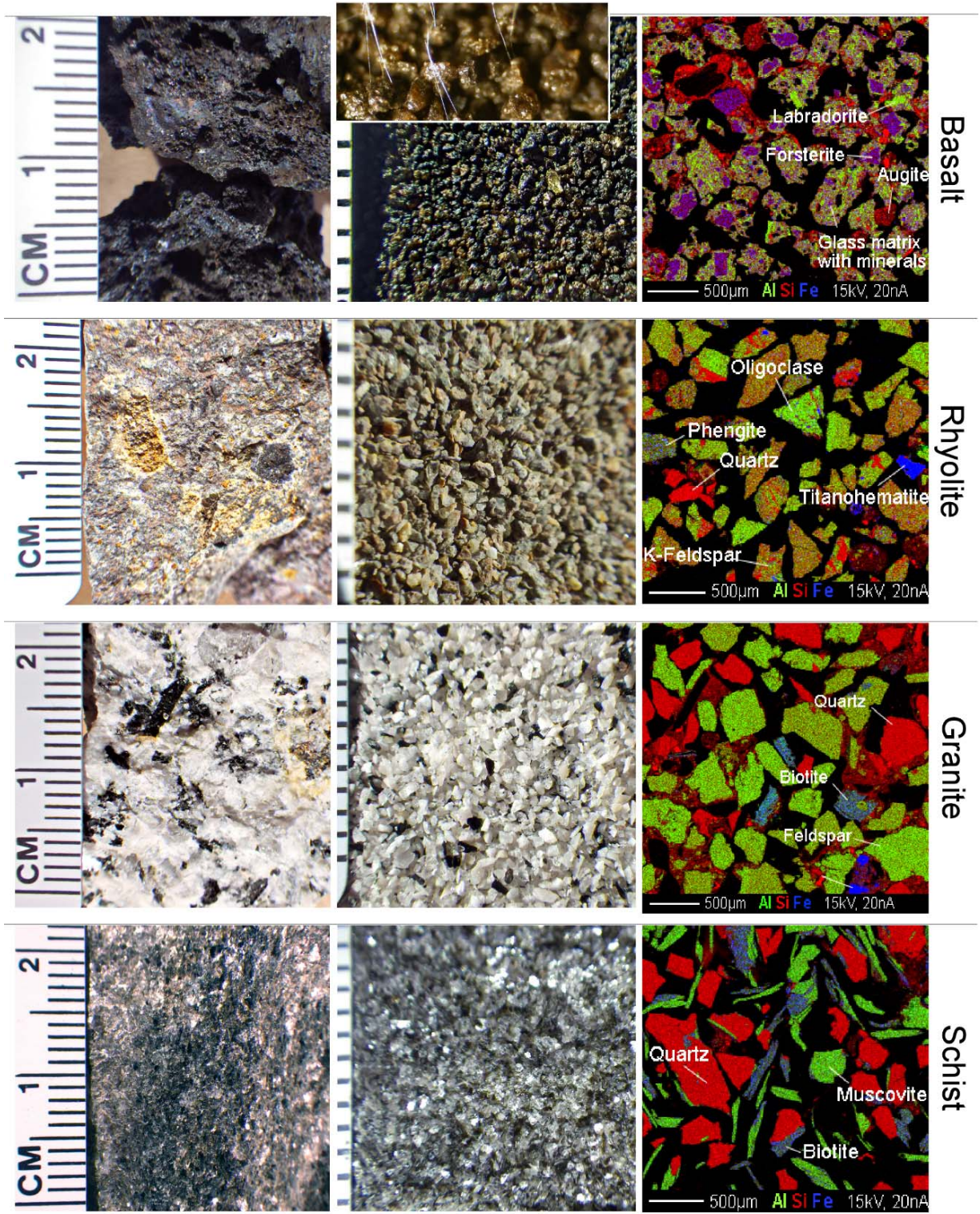

*Figure 1* **Physical and mineralogical differences among unreacted substrates.** Physical features of the unprocessed (left) and processed (center, scale in mm) rocks used as experimental substrates. Right, electron microprobe X-ray reflectance images showing elemental distribution in the granular material. Inset in basalt (center) is a stereomicroscope image depicting 1µm-thick mycorrhiza hyphae connecting plant roots (not imaged) to rock particles.

imprinted a foliar texture, and exhibited the largest pore volume-to-density ratio among substrates. The molar abundance of major elements in the unreacted substrates (and their summed % oxides - indicating their total contribution to rock matrix) followed the trends (also Fig. s2A):

Basalt: Si > Al > Mg > Ca > Fe > Na > Ti > K > P > Mn; 99.82±02 %
Rhyolite: Si > Al > Na > K > Fe > Ca > Mg > Ti > P > Mn; 99.03±0.81 %
Granite: Si > Al > K > Fe > Na > Ca > Mg > Ti > Mn > P; 99.85±0.78 %
Schist: Si > Al > K > Fe > Mg > Ti > Na > Ca > Mn > P; 97.54±0.72 %

The geochemical compositions of unreacted substrates were generally consistent with known deviations by lithology relative to mean upper continental crust (Fig. s2B). Basalt was enriched in Mg, P, Ca, Ti, Mn, and Fe as compared to the crust and the other rocks, providing better biotic colonization opportunities. Rhyolite and granite were depleted in Mg, Mn, and Fe, whereas schist contained somewhat lower concentrations of Na, P, and Ca (Fig. s2). The distinct physical and chemical properties were expected to affect not only abiotic denudation, but also plant and microbial biofilm establishment, biomass accumulation, and element stoichiometry in the evolving rock-biota system. In turn, compared to the abiotic control (C), biotic metabolism was expected to affect element stoichiometry and mineral weathering rates according to biological requirements.

**System evolution through time.** Over the experimental window of 20 months, effluent water pH decreased by about 0.5 - 1 log units in all substrates except granite, with greatest shift in the abiotic treatment in basalt and schist (Fig. s3A). Electrical conductivity — a synthetic measure of solute production during chemical denudation — decreased sharply during the first two months, and it reached the lowest, steady phase after about 300 weathering days.

With few exceptions, fractional removal of rock-derived cations increased sharply during the first two months, after which a number of element-specific patterns developed (Fig. 2). Removal of Si, Al, K, and Mg increased slowly but steadily, consistent with the dissolution the rock matrix. Phosphorus ( a limiting nutrient) and Mn (a micronutrient) in granite and schist leveled off after an initial 60-120 days of leaching (while it continued to increase in basalt and rhyolite). Similar trends were observed for anions, with a sharp increase indicating rapid loss in total dissolved forms during the first 30-60 days, followed by a plateau (greatly diminished loss) for most species, except fluoride and phosphate in basalt and rhyolite which continued to increase (Fig. s4A). Where present, a biotic signal started to differentiate treatments during the most active weathering period (first 60 days) in both leached cations and anions, plateauing thereafter (Fig. 2 and Fig. s4). Specifically, the clearest positive biotic signals were for Mg and Ca in basalt, K in rhyolite, and negative for Cl in rhyolite and Al, Fe, Mn, and Cl in schist.

The largest accretion in plant biomass (above and below ground) occurred during the first four months (with the largest occurring in rhyolite, Fig. s5A, B), implying accelerated plant development (nutrient uptake) during the period of most rapid nutrient release, followed by little-to-no change. This suggests a biomass conservation strategy during the inception of nutrient limitation following initial column exploration, clearly supporting a previous model which hypothesized a shift between high geosphere nutrient output (weathering) and high nutrient uptake and recycling (limitation) of stable substrates over longer ecosystem development times[23,24], although atmospheric N limitation should also be expected in natural settings. This is also supported by significantly larger energy investment in root development as compared to shoot (Fig. s1C; ref.[20]). The biomass in schist followed a slow growth over time. Water consumption, reflecting ecosystem metabolism (mostly evapotranspiration), slowly increased over the experiment, and a biotic effect (increased retention in subsurface environment) was largest in rhyolite (B *vs.* C) and schist (BG *vs.* B) (Fig. *s5C*), highlighting the potentially important role of these substrates in water retention and ecosystem development in natural landscapes. Mycorrhiza contributed significantly to plant biomass accumulation in rhyolite (Fig. *s5B*), and subsurface water retention (decreased evapotranspiration) in basalt and schist (Fig. *s5C*), consistent with previous suggestions that mycorrhizae provide access to water in mineral pore spaces and help retain water in soil environment[25].

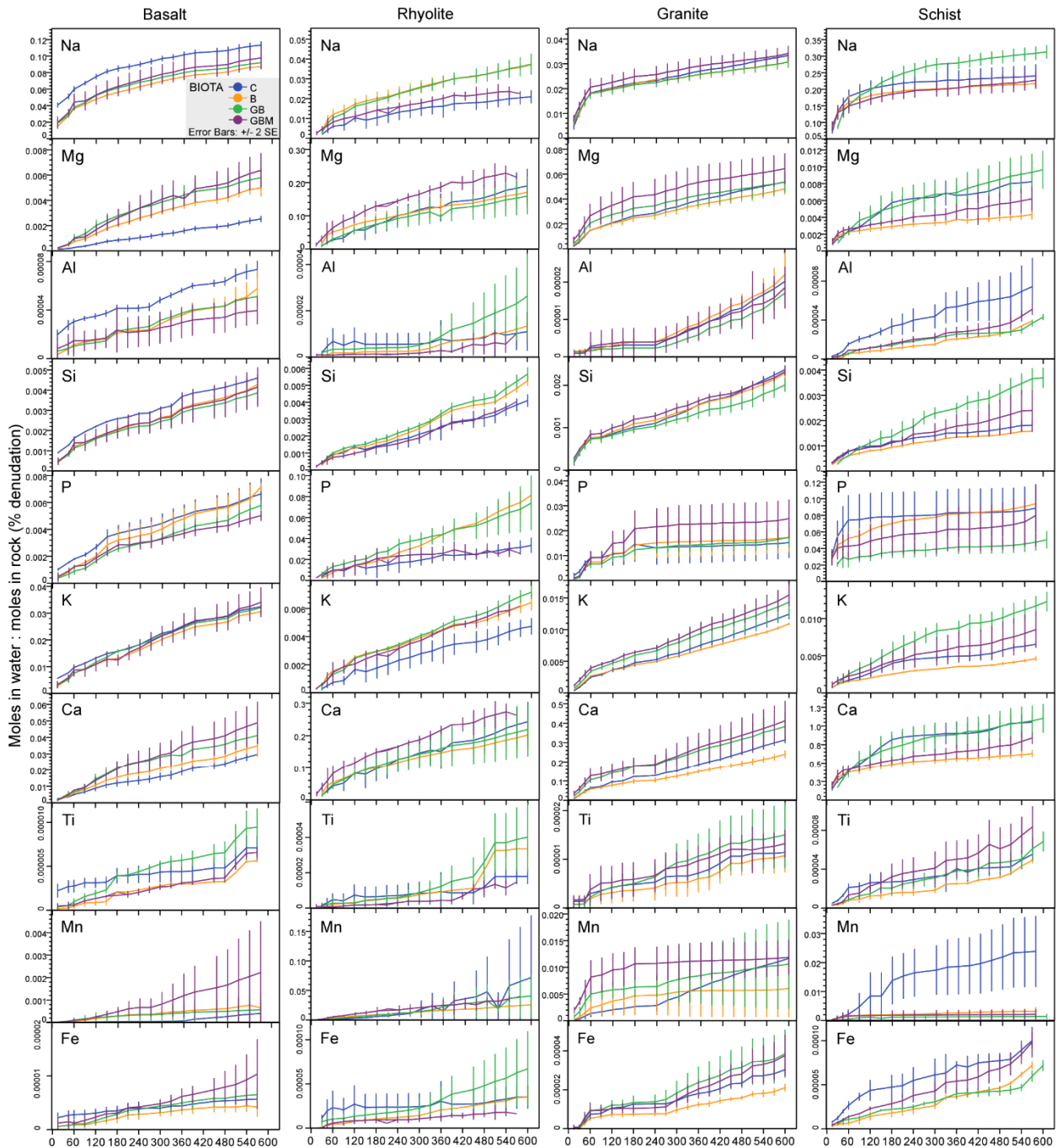

*Figure 2* **Cumulative fractional removal in the dissolved load.** Time-lapse trends of cation fractional leaching from rock substrate in pore water over the two-year weathering experiment (X axis in days). Values represent cumulative column means of total element loss in pore water per sampling event as a fraction of the total (molar) mass of a given element present in rock at the beginning of each experiments. Treatments are: (C) control, (B) bacteria, (BG) bacteria - grass, (BGM) bacteria - grass - mycorrhiza. Error bars (+/- 2SE of column triplicate) and color-coding are for all plots. N = 60 sampling events.

The early treatment effects on element mobilization (Fig. 2 and Fig. *s4*) reflected rapid biotic modulation of incipient oxidative weathering, which largely depended on the physical and geochemical properties of the substrate. The observed initial pulse in solute output may be due, in part, to preferential dissolution of fine particles that could have adhered to larger grain surfaces despite attempts to develop a particle size-controlled substrate. However, it can also be attributed to the fact that the substrates all comprised freshly exposed mineral surfaces and micro-fractures could have developed during incipient interaction with water (hydration) and air (oxidation). Electron microprobe analysis of rock grains (abiotic control) before and after the experiment, compared with field-weathered basalt collected from the same material excavation site, further hint at an incipient physical effect (e.g., cracking developed during oxidation/hydration of mineral surfaces; Fig. *s4B*). Evidence of micron-scale surface spalling, associated with minimal secondary mineral deposition has been reported for subsurface basalts in the field[26], and for laboratory experiments (repulsive forces during water-rock interaction)[27], indicating an incipient physical effect. However, secondary mineral nucleation (and subsequent crystal growth by ion sequestration), if present, should also have occurred during early, high relative saturation phase of the experiment, as the activation energy barrier for critical nuclei formation is more easily met.

Surprisingly, aqueous speciation and saturation index (SI) calculations with respect to potential secondary mineral phases by Visual Minteq (J. P. Gustafsson; https://vminteq.lwr.kth.se/) identified conditions for secondary mineral formation consistent across substrates. Effluents of all igneous rocks indicated supersaturation with respect to Fe, Mn, and Mg oxides (hematite, bixbyite, magnesioferrite) and Ca-phosphate (carbonate fluorapatite - CFA and hydroxyapatite in all but granite) at SI > 10. To a lesser extent (5 < SI < 10), they also favored formation of Fe (hydr)oxides (maghemite, goethite, lepidocrocite, ferrihydrite) and the secondary aluminosilicates kaolinite, halloysite, and imogolite (less so in granite), and also hydroxyapatite (in granite). Effluent solutions from metamorphic schist indicated favorability of precipitation of Fe and Mn-Fe oxides (bixbyite, hematite; SI > 10), and CFA, Fe and Mg (hydro)oxides (maghemite, goethite, lepidocrocite, magnesioferrite), and kaolinite (5 < SI < 10). Similarly, conditions favored dissolved organic matter-ion complexation, particularly with Ca, in all rocks. However, a lack of observed secondary mineral formation on grain surfaces during electron microprobe analysis (Fig. *s4B*) suggests kinetic limitation of secondary phase precipitation. Further analysis of fine particles (not shown here) and sequentially extracted fractions should shed more light on potential secondary mineral formation during incipient weathering.

**Substrate controls on solute export.** Total dissolved cation loss by rock type in the abiotic control followed the trend granite > basalt > rhyolite > schist, while total anion loss decreased in the order granite > schist > basalt > rhyolite (Table *s1*). The cationic trend was consistent with a parallel decrease in electrical conductivity (a synthetic measure of ionic strength) and an increase in proton export (lower pH), consistent with an abiotic co-dependence of solid dissolution and pore water acidity. The loss of major cations in effluent solutions (total µmoles) from the abiotic control (and their abundance in initial rock) over the 20-month experiment followed the order:

Basalt: Na > Ca > Si > Mg > K > P > Al > Mn > Fe > Ti (Si > Al > Mg > Ca > Fe > Na > Ti > K > P > Mn);
Rhyolite: Si > Ca > Na > Mg > K > Mn > P > Al > Fe > Ti (Si > Al > Na > K > Fe > Ca > Mg > Ti > P > Mn);
Granite: Ca > Na > Si > K > Mg > P > Mn > Al > Fe > Ti (Si > Al > K > Fe > Na > Ca > Mg > Ti > Mn > P);
Schist: Ca > Na > Si > K > Mg > P > Al > Mn > Fe > Ti (Si > Al > K > Fe > Mg > Ti > Na > Ca > Mn > P).

According to above results, all substrates leached comparatively more Na, Ca, and Si than other major cations (regardless of their rock abundances), due to limited precipitation of Na and Ca – bearing solids and high abundance of Si. Fractional release order (quantified by the total moles released in solution divided by the total moles in rock) showed that Fe, Ti, and Al leached an order of magnitude less than the rest of structural elements from all substrates (Fig. 3). Basalt preferentially exported K, rhyolite preferentially exported Mg and Mn, and schist preferentially exported Ca, Na, P, Fe, and Ti (Fig. 3A), which should have further impacted biological uptake. Abundances of dissolved organic and inorganic C and total N increased initially, then plateaued and showed rock-specific biological effect (Fig. *s3B*).

Multivariate analysis of pore water solute concentrations in abiotic controls identified element groups with similar dissolution patterns (Table *s2*). Most elements in basalt grouped together (PC1), which is consistent with dissolution of glass, driven primarily by hydrolysis and carbonation reactions (70.2% of total variance in pore water element content, Table *s2*). The clustering of Group II elements Ca and Mg (PC2) is consistent with their similar electronegativity (1.36 and 1.31, respectively) controlling their co-dissolution. Carbonation and hydrolysis reactions also dominated in rhyolite (78.6% of total variance; PC1, 2 and 3) and schist (57.6% PC1 and PC2, when excluded TOC), while in granite the influence of TOC ligands was greater (PC1, 38% of total variance; Table *s2*). Organic and inorganic forms of C and N have been documented in igneous terrestrial and lunar rocks before[28–30], and they were also present in our unreacted materials (the following values in µg kg$^{-1}$: basalt TOC = 0.029, TIC = 0.037, TN = 0.00085; rhyolite TOC = 0.039, TIC = 0.055, TN = 0.0087; granite TOC = 0.01, TIC = 0.12, TN = 0.002; and schist TOC = 0.0091, TIC = 0.017, TN = 0.0014), although the source of the organic fraction in the un-inoculated material is still unclear. It therefore appears that in 'mineral-only' weathering environments, on the early Earth, in present day uplift areas of intense incipient weathering (e.g. volcanoes and mountain tops), and perhaps on Mars, geochemical cycles of major bedrock elements would be dominated by carbonation of highly soluble silicate-rock constituents with important contribution (most likely through oxidation) of redox-sensitive elements. These abiotic drivers of weathering should also underlie further biological effects in the wider terrestrial biosphere.

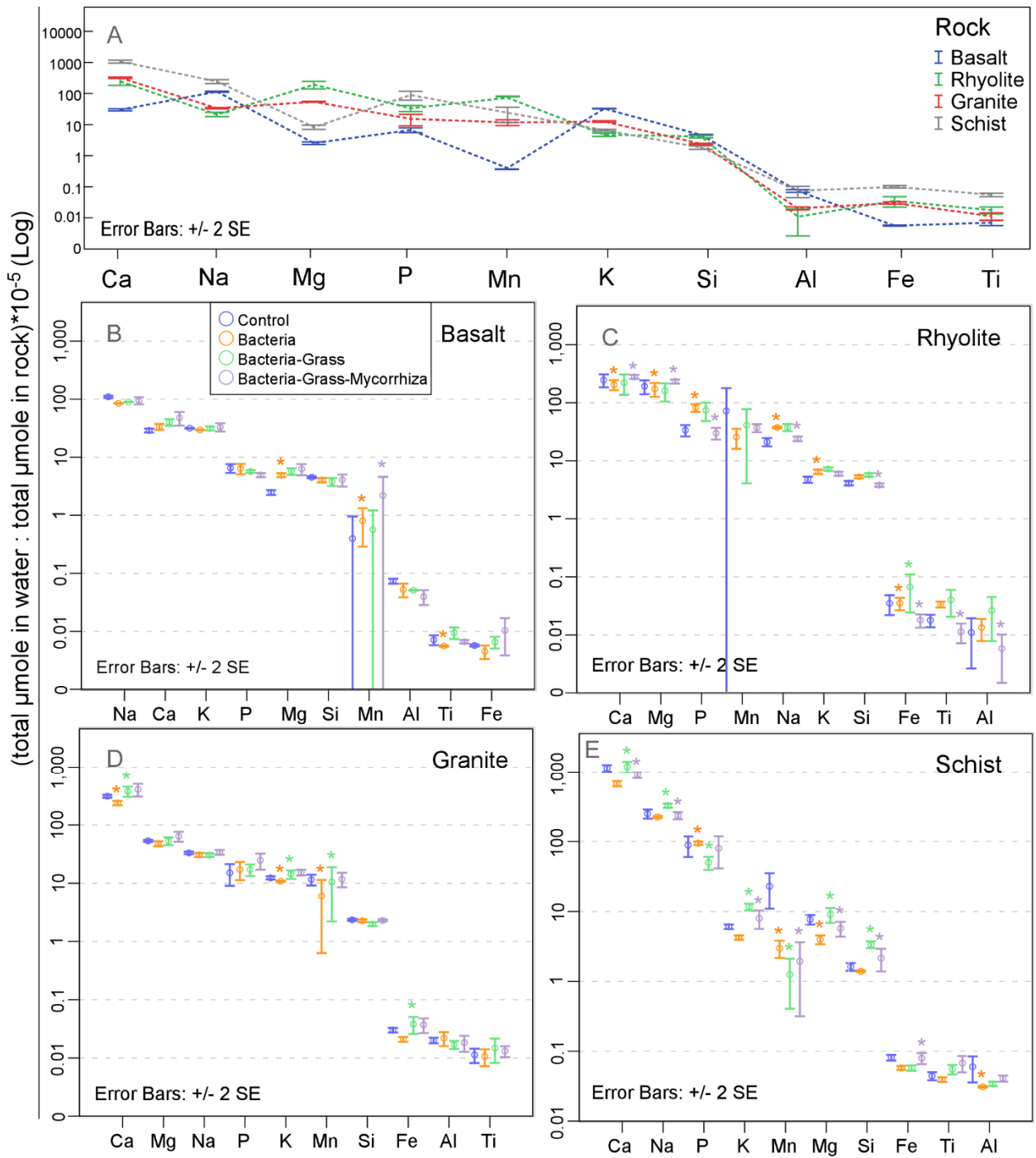

*Figure 3* **Substrate and biotic effects on cation export in pore water.** (A) Rock-normalized total denudation (preferential leaching) of major cations in pore water in un-inoculated control, summed across the 60 sampling events, and (B-E) across biological treatments in the 4 tested rocks. Means and standard errors are calculated from column triplicates. Excessive Mn SE bars are due to one triplicate displaying consistently high values throughout the sampling period. Elements are arranged in order of decreasing values.

**Biotic controls on element export.** Through interaction with the abiotic substrate, microbes, plants and associated mycorrhiza are expected to change element stoichiometry through individual and synergistic effects, which can potentially be captured in pore water chemistry. Electric conductivity (EC), a synthetic measure of total ion concentration, generally increased from chemical-only to BGM dominated systems in basalt (48.5%), rhyolite (17.9%) and granite (26.5%), but decreased in schist (-32.9%), coinciding with similar trends in cation and total nitrogen (TN) export (Table s1). Under the same treatment conditions, column retention of water (less evapotranspiration) and total anions (less water loss) increased in all rocks. Values of pH were decreased in both control and biotic treatments according to: basalt > granite > rhyolite > schist (Table s1). A reduction in pH inter-substrate variation coefficient from 7.6% (control) to 3.8% (biota), hints to a buffering capacity of life when colonizing various abiotic substrates, bringing them to a pH range more suitable to growth.

The introduction of an active microbial community increased Mg and Mn export from basalt, Na, P, K and F from rhyolite, and P from schist, relative to the un-inoculated control (Fig. 3B-E, Fig. s6). Conversely, it decreased aqueous export of Ti in basalt, Mg, Ca, and Fe in rhyolite, K, Ca, and Mn in granite, and Mg, Al, and Mn in schist. The addition of vascular plants increased (relative to microbes only) Fe export in rhyolite, K, Ca, Mn, and Fe in granite, and Na, Mg, Si, K, Ca, and F in schist, while decreasing Cl release from basalt, S, Cl, and Br from rhyolite, S from granite, and P, Mn, Cl, and Br in schist. Arbuscular mycorrhiza stimulated Mn loss in pore water in basalt, Mg and Ca in rhyolite, and Mn and Fe in schist. Conversely, Na, Al, Si, P, Ti, Fe, and F in rhyolite, and Na, Mg, Si, K, Ca, and F in schist were lost in smaller amounts under the effect of mycorrhiza (Fig. 3B-E, Fig. s6). These results support the hypothesis that arbuscular mycorrhiza are active components of the terrestrial weathering engine, selectively influencing element cycles[31]. Manganese, a redox-sensitive element, displayed consistently high values in some column effluents, which resulted in high triplicate variability. Similarly, nitrite and nitrate, displayed a peak in leaching during the first 2 months and close to 0 values after that (Fig. s4), explaining its high average variability (Fig. s6).

Analysis of ion co-dissolution by treatment revealed that $Na^+$ and $SO_4^{2-}$ - two highly soluble ions - generally grouped together in the major PCs (interpreted as major element sources in minerals and dominant dissolution process; Table s3); however, a nuanced biotic-specific pattern emerged. Microbial inoculation modified P and Br PC group membership (Table s3) compared to the abiotic control (Table s2) in all substrates but schist (PC1), and nitrite membership in all substrates but granite. Grass affected Mg (a component of photosynthesis) group membership in all rocks (PC2) and phosphate, essential for cell division and energetics, in all rocks but schist (PC3 and PC4), likely due to their increased uptake. The mycorrhiza effect was mostly rock-specific (Table s3). While only exploratory in nature, PCA results indicate an important role of ecosystem's biological constituents in modifying the inter-elemental relationships during their mobilization in ways different than the abiotic background. This clearly supports a previous hypothesis that biological stoichiometry drives changes in mineral weathering stoichiometry[23], providing the potential for *in situ* biosignatures.

In analyzing the likely drivers of element leaching in the system, we observed that carbonation and its nested variables $H_2CO_3$, bicarbonate ($HCO_3^-$), carbonate ($CO_3^{-2}$), pH ($H^+$), appear to dominate dissolution in all biotic treatments and rocks, except granite (% variance explained by predictors for each PC, Table *s3*, summarized in Fig. *s7*). This reflects a $H_2CO_3$-dominated low-organic weathering system under incipient ecosystem colonization. Bicarbonate, the first anion in the dissolved $CO_2$ metabolism (resulting from $H_2CO_3$ dissociation), was the dominant cation charge balancing anion, and it explained the largest variation of the element groups (PC1). Its influence in the system increased from un-inoculated to microbial treatments in all rocks, clearly supporting a buffering capacity of pore environment by microbiota, further enhancing element dissolution. Under grass, bicarbonate influence on element leaching increased over microbes in basalt and rhyolite and decreased in granite and schist, while under mycorrhizae it decreased with respect to grass in all rocks but schist (Table *s3*, Fig. *s7*). Moreover there was a significantly (ANOVA *p*<0.05) higher total moles of bicarbonate released in the biotic versus abiotic systems, which is important to mention, since bicarbonate is an often used measure of silicate weathering by carbonation reactions[32]. Carbonate, an indicator of further alkalinization had the opposite effect to bicarbonate, i.e., decreasing its influence on element leaching from abiotic to added ecosystem complexity, and this was clearly visible in basalt (Fig. *s7*). Overall, carbonic acid had a greater influence in the biotic system than control. The influence of pH was generally greater in the abiotic control, further supporting a buffering of weathering environment by biota. In granite, the role of organic C in element complexation appeared to be greater than the inorganic pathway.

Taken together, these results show that in an oxic weathering environment, the systematic introduction of different ecosystem components produces differences in incongruent leaching of major rock constituents, despite the stabilizing effects of proton concentrations. This is largely rock-specific and seems to depend on physiological needs of different life forms. However, carbonic acid appears to drive incipient bioweathering across substrates, a process which is enhanced by biotic respiration.

**Plant uptake and the effect of arbuscular mycorrhiza.** Appropriate substrate chemistry (*e.g.* pH and solute availability) is essential for biotic establishment and can dictate element cycles through the developing ecosystem. Plant-essential nutrients such as P, K and Mg decreased substantially in biomass over the growth period (Fig. *s8*), suggesting nutrient limitation. This was also supported by lower element abundances in tissues with increasing biomass, particularly N, Al, Ti in basalt, N, P, K, Mg in granite, N, P, K in schist, and K in rhyolite (Fig. *s9*). Shoots appeared to more directly reflect nutrient limitation, in line with previous findings on Mg deficiency in *Arabidopsis thaliana*[33].

Phosphorus, C, and N, three critical elements supporting life[34], are required in stoichiometric balance for the maintenance of body mass, protein and enzyme architecture, energy fluxes (ADP-ATP), and, as part of DNA and RNA, for cell physiology and reproduction. While C can easily be acquired through photosynthesis, N is an "expensive" nutrient fixed by oxygen-sensitive nitrogenase in free-living and symbiotic microbes, with relatively smaller amounts originating in the weathering substrate itself[30]. Phosphorus, Mg, and K as well as the other major elements are strictly limited by availability and solubility of the requisite mineral phases. Magnesium, forming 0.2-0.4% of dry biomass (mostly part of chlorophyll) is strictly necessary for plant growth, and is a necessary activator for many critical enzymes, including ribulosbiphosphate carboxylase (RuBisCO) - and phosphoenolpyruvate carboxylase (PEPC), both essential in $CO_2$ fixation[35,36]. Potassium regulates plant growth through protein synthesis, and maintaining cytosol turgescence by anion/cation balance[37,38]. Despite the possibility of major nutrient limitation (N:P:K:Mg), plants grown in our system managed the available resource in ways that sustained their development and stimulated mineral dissolution, albeit at the cost of small body mass (Fig. s5A and B).

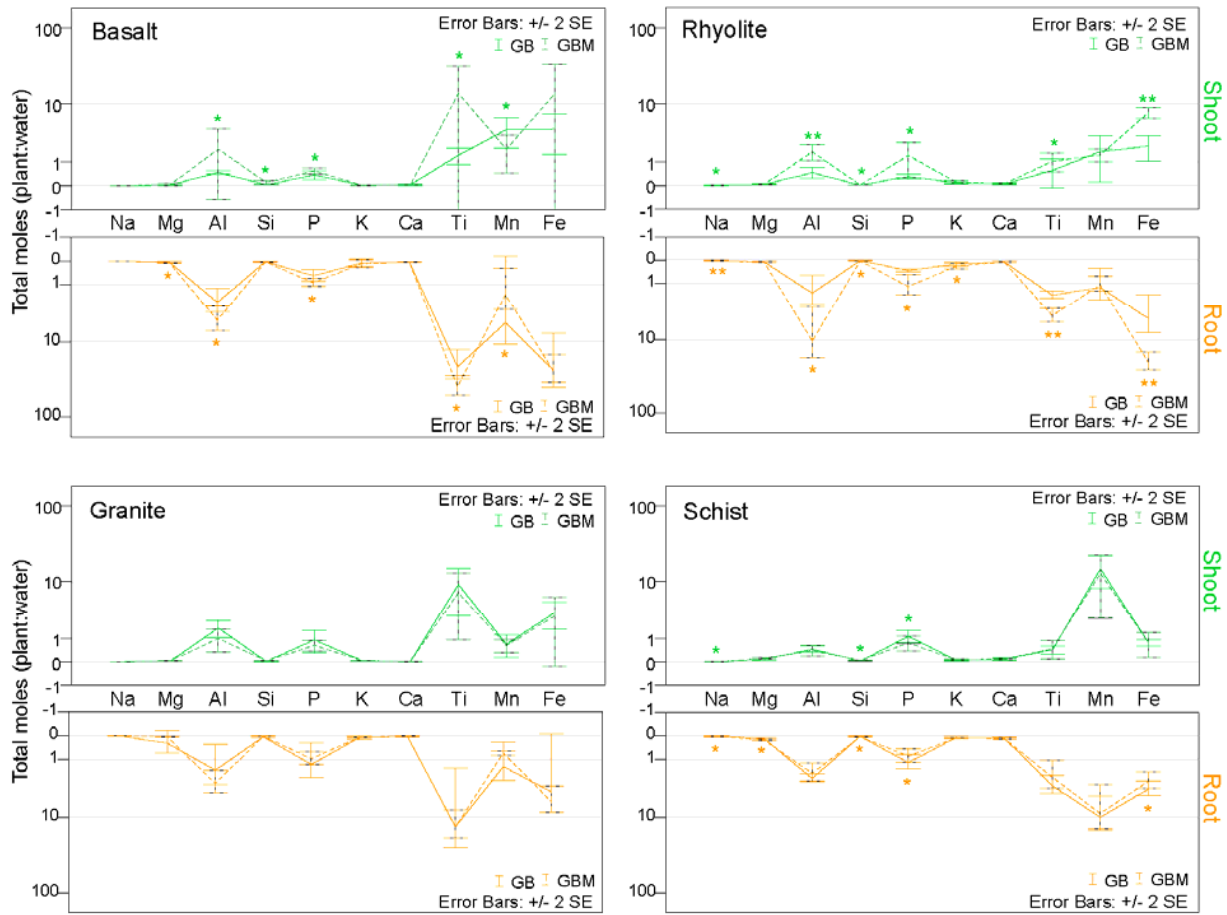

*Figure 4.* **Preferential plant uptake.** Water-normalized element uptake in plant (root and shoot) with (GBM) and without (GB) arbuscular mycorrhiza, on different rock substrates. Treatment effect is significant at **+/-2SE and *+/-1SE levels.

Aluminum, P, Ti, and Fe all exhibited high fractional partitioning from water to tissue, due to their lower dissolved abundances (Fig. 4). Mycorrhiza clearly stimulated the preferential uptake and transfer to shoot of P and the transfer of Si in basalt and rhyolite, and inhibited them in schist, which has interesting evolutionary implications as same fungal species seems to reflect either, mutualism or competition (parasitism) with the same host when substrate changes[14]. Mycorrhizae also stimulated the uptake of other elements including Al (toxic at high concentrations) and Ti in basalt and rhyolite, and Na transfer to shoot in rhyolite and schist, while decreasing Mg plant uptake in basalt and Na in rhyolite and schist. It also inhibited the uptake and transfer of Mn in basalt (Fig. 4). Overall, the element concentration results suggest a tight interplay between fungal physiological needs, plant physiological needs, mineral dissolution rates, and nutrient availability in the granular substrate. This results in an intricate elemental compartmentalization pattern, even though the system was designed to greatly reduce the complexity of the natural environment.

**Formation and preservation of *in situ* biosignatures.** Incongruent mineral dissolution modulated by abiotic and biotic processes should leave identifiable geochemical traces in weathering byproducts[39]. To help distinguish the effect of different ecosystem components on incongruent element mobilization, and hence to establish a biological fingerprint, we propose a biological signature index, BSI (E. *1*), with $x$ detailed in E. *2*. The numerator of $x$ represents the moles element ($i$) to moles total cations {Na,…,Fe} in the extracted fraction ($f$), while the denominator represents the moles of element to moles total cations in rock ($r$). $Z$ is the abiotic fraction of $x$ (0-100 scaled), and it is rock specific (Table *s4*).

The index is unitless, can take values from -100 (negative values for less extracted element fraction compared to control, 0) to 100 (positive values for more extracted fractions compared control, 0).

$$BSI = 100\frac{x-min(x)}{max(x)-min(x)} - Z \text{, where}$$

(E. *1*)

$$x = \frac{\frac{i_f}{\sum_{i_f=Na}^{Fe} i_f}}{\frac{i_r}{\sum_{i_r=Na}^{Fe} i_r}}$$

(E. *2*)

BSI distributions for water, exchangeable and poorly crystalline fractions, were rock-specific (Fig. 5). BSI revealed a well-differentiated microbial fingerprint in water, (both for incongruent leaching and column retention as shown by values above and below the abiotic reference line in Table *s4*) for all rocks, mostly for group IA and IIA elements of relatively low electronegativity and high solubility (Ca, Mg, Na), which is a strong signature of biologically catalyzed weathering (Fig. 5). Vascular plant growth had a pronounced fingerprint in all rocks but rhyolite, mostly from Si, an element of strong electronegativity that is vital for cell walls. The presence of mycorrhizae had the strongest stoichiometric signature in water in rhyolite; however, it affected less soluble transitional metals, such as Ti across rocks.

The exchangeable fraction, which comprises weakly bound bioaccessible elements, contained less apparent biosignatures than the dissolved fraction; however, a strong microbial fingerprint developed in rhyolite for group IA- IIIA elements and within the redox-sensitive Fe-Mn group. Biosignatures due to grass growth in this fraction were only present in schist suggesting a weathering system still in its early phases. A fungal biosignature was strong in rhyolite and schist, particularly for Ca, K, and Al (Fig. 5).

The more stable, poorly-crystalline fraction generally stored less obvious biosignatures than the exchangeable and dissolved fractions, which is unsurprising since a higher signal/noise is expected for incipient weathering. However, it had a marked microbial and fungal fingerprint in rhyolite, and plant in schist. Phosphorus, a critical element for all components of the biosphere, showed elevated biosignature index values in all biological treatments (microbes, grass and mycorrhiza) mostly in rhyolite and schist, and in all fractions under microbial presence only in rhyolite. From our results, it is clear that as weathering initiates, the biological fingerprint is strongest in leached pore water, and it decreases as elements bind to mineral and mineral-organic surfaces, then nucleate in poorly crystalline fractions, with rhyolite and schist better integrating information of life's presence. This is most likely because major elements enter more and more complex organic and inorganic functions, which will ultimately dictate ecosystem development in the landscape and major biosphere cycles on longer timescales.

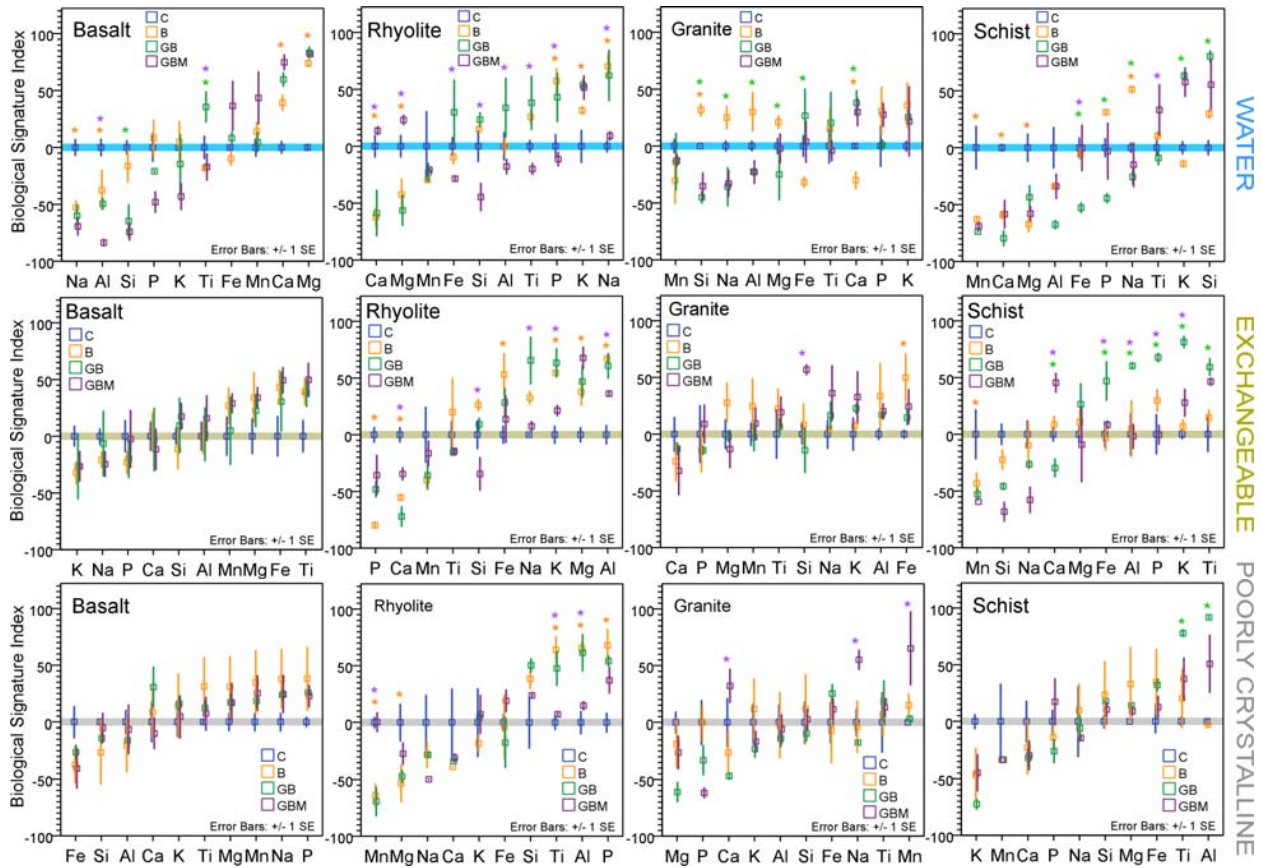

*Figure 5.* **Formation and preservation of biological signatures.** Effect of different ecosystem components on incongruent leaching and retention of elements in water (top), exchangeable (middle) and poorly crystalline (bottom) fractions, at the end of the two – year experiment, as measured by the Biological Signature Index (E *s.1*). Horizontal line is set at abiotic control. Treatment effect is significant at 99.5% level (Fisher's least significant difference, ANOVA) for *B *vs.* C, *GB *vs.* B, *GBM *vs.* GB comparisons.

**Mass balance analysis.** Upon release from rock, elements were partitioned into one of five biogeochemical compartments: dissolved, plant tissue (roots and shoots), surface-adsorbed (targeted here by ammonium acetate extraction, AAE), and secondary minerals (the poorly-crystalline fraction of which was targeted here by acid ammonium oxalate extraction, AOE). Generally, extracts of the solid-phase, particularly the poorly-crystalline fraction, dominated over the total mobilized (water and biomass) pools by an order of magnitude (Fig. 6 and 7), even though their biosignatures were weak. The fractional distribution of most abundant rock constituents (Si, Mg, Na, K, and Ca) were higher in the solid phase extracts, followed by water and plant biomass. Calcium in all rocks, Na and Mg in rhyolite, Na, Mg, and Mn in granite, and Mg and Mn in schist were similar or more abundant in the exchangeable than the poorly-crystalline fraction, reflecting their enhanced bioavailability. Interestingly, Fe and Mn were retained significantly in the solid phase (extractable in AOE) over other fractions, illustrating their oxidation (and precipitation from solution) to Fe(III) and Mn(IV) from Fe(II) and Mn(II) found in rock. Phosphorus, Al, Ti, Mn, and Fe, the least abundant of major elements in rock, were generally more abundant in plant tissue than in water. Root tissue concentrations exceeded those of shoots, representing a vital energy investment in the nutrient-poor environment. This is important as P is a critical element for life, and high absorption affinity from relatively low environmental abundances has significant consequences across the biosphere. The four elements are also expected to be significantly constrained by biosphere during their planetary cycles.

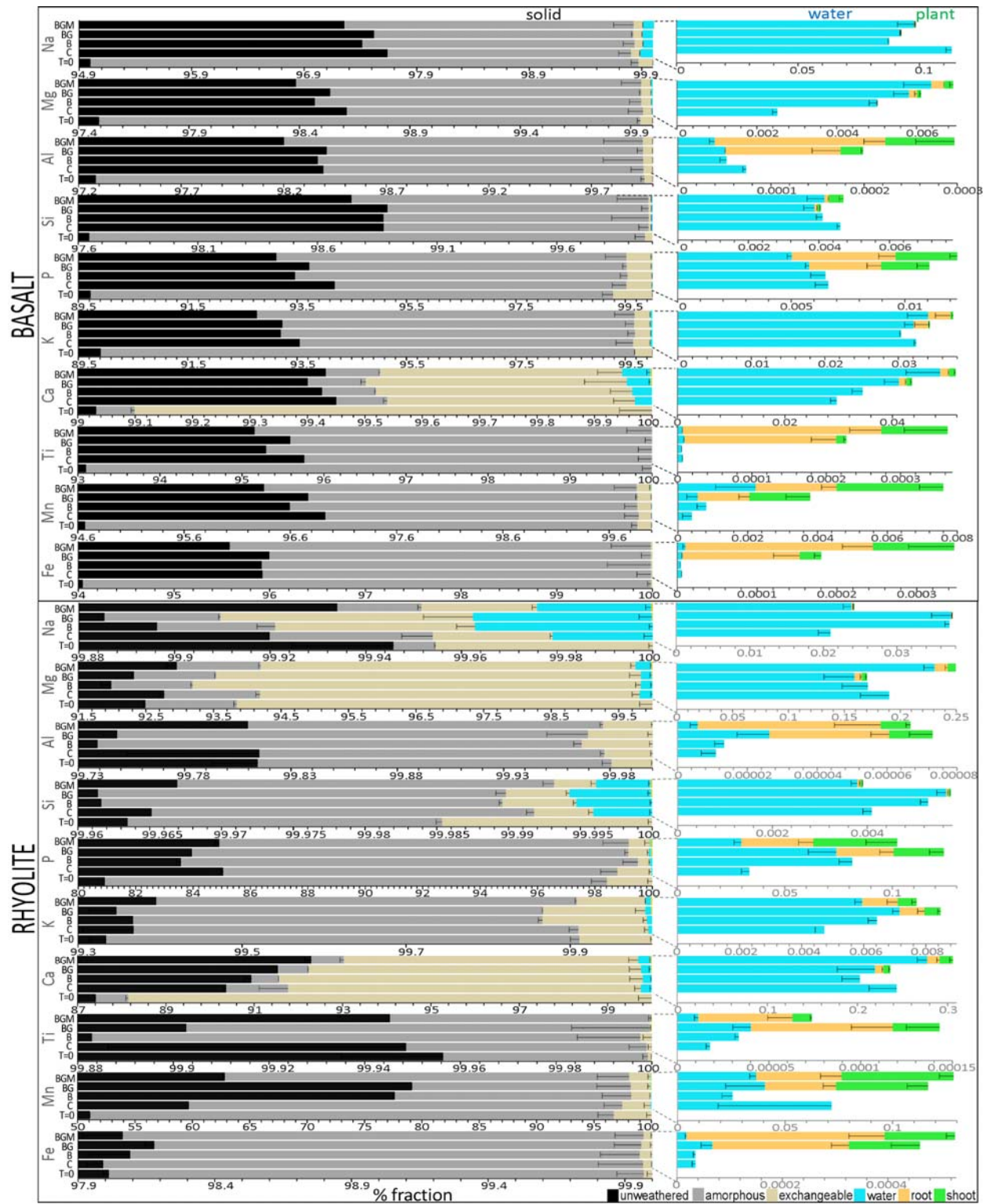

**Figure 6 Fractional compartmentalization of biological weathering products in basalt and rhyolite.** Cation distribution in biomass, aqueous, exchangeable, poorly crystalline, and residual pools across the four biotic treatments after two years of biological weathering. Aqueous values are means of treatment triplicates summed across the 60 sampling events. Solid fraction and biota were analyzed at the end of the experiment. T=0, initial rock. Treatments: C, control; B, microbes; BG, microbes-grass; BGM, microbes-grass-arbuscular mycorrhyza.

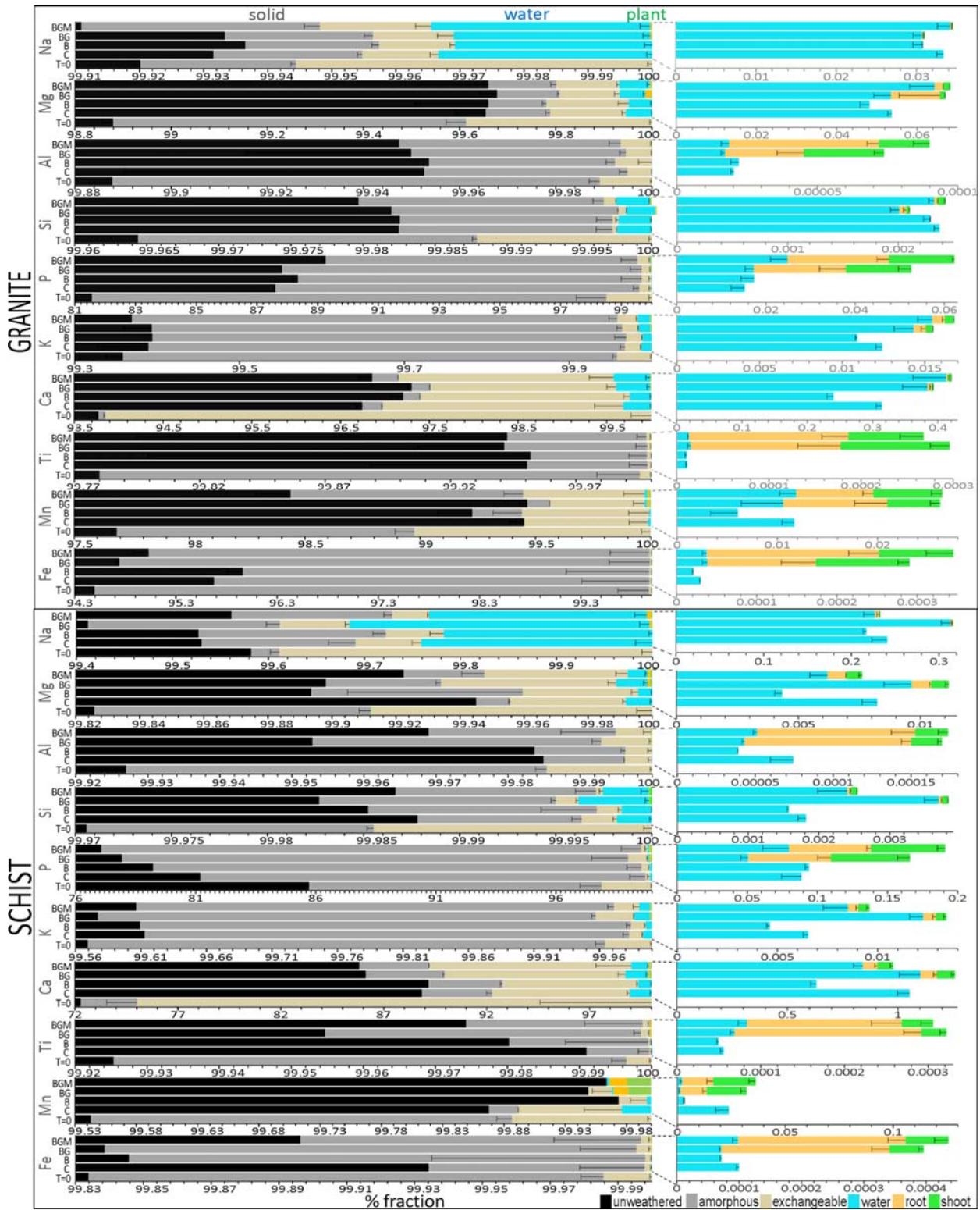

*Figure 7* **Fractional compartmentalization of biological weathering products in granite and schist.** Percent fractionation of major rock cations in liquid, solid and biological phases among the four biotic treatments. Water values represent means of treatment triplicates summed across the 60 sampling events (603 days experimental period). Solid fraction and biota were analyzed at the end of the experiment. C, control; B, microbes; BG, microbes-grass; BGM, microbes-grass-arbuscular mycorrhyza; T=0, initial rock.

Compared to the abiotic control, microbes increased weathering (i.e., extractable plus dissolved plus biomass) in basalt and schist by 7.5% and 5.5%, respectively and decreased it in rhyolite and granite (by 21.6% and 5.9%; overall decrease across all rocks of 8.5%), while also decreasing total ion mobilization in water by 23.4% across rocks. This implies that microbes inhibited weathering by coating active mineral surfaces and retaining elements in biofilms and/or secondary minerals, potentially by stimulating (either by active or passive mechanisms) the nucleation of secondary minerals[40]. The retained element pool could be further used by other ecosystem components, e.g. plants and protozoans, once primary minerals are weathered[9]. However, microbes decreased total weathering of Fe and Mn in rhyolite, Mn in schist (potentially by forming biogenic manganese oxides, e.g. birnesite[41], not extracted by AOE) and Ca in rhyolite and schist, and stimulated leaching of Mg, Ca, and Mn in basalt, Na, K, Si, and Ti in rhyolite, P in rhyolite, granite and schist, and Al in rhyolite and granite.

Plant colonization increased total mobilization (water + plant) of all elements in the four rocks by a factor of 10 in many cases (overall mobilization across rocks of 63.4%), with largest effect in schist. This enhanced mobilization was associated with an increased extractable pool and total weathering only in granite (exception: Mg, Ca, and Mn) and schist (exception: Mg). In contrast, plants decreased total weathering (water + plant + exchangeable + poorly crystalline) of all elements but Ca in basalt, and most elements in rhyolite, and this was due to general decrease in solid fractions (AAE and AOE). These substrate-dependent effects on weathering would have had a transformative impact on Earth surface environments and global biogeochemical cycles when vascular plants colonized the land during the early Phanerozoic, the overall effects of which have only been partially explored[42].

Compared to the effect of plants, fungi (where present) reduced overall total leaching across all rocks by about 7%, basalt being a remarkable exception where fungi stimulated leaching, total mobilization, and total weathering of most cations (consistent with increased Ca leaching from basalt in mature forests[31]), and this was also related to increased size of solid pools. With few exceptions (Fe, Mn, P, and Ca), mycorrhizae diminished weathering and mobilization in rhyolite and schist, where solid pools were also lower than in grass-microbes. This rock variability in allocation of weathering pools (i.e., compartmentalization) under abiotic and biotic treatments suggests that the three major ecosystem components, i.e., microbial communities, vascular plants, and mycorrhizae, all played crucial but differential roles in element redistribution during various epochs of land colonization by an expanding biosphere.

It is noteworthy that in the unreacted rock, the solid pools (targeting exchangeable and poorly-crystalline secondary minerals) were larger than in the abiotic and biotic treatments, particularly in basalt (Fig. 6, Fig. 7). While this was initially surprising, it is most likely related to the higher reactivity of fresh crystal faces, fine particles and microfractures resulting from rock grinding, which are susceptible to rapid dissolution in AAE and AOE.

**Contribution of the biosphere to global weathering**. Considering our leaching data for the four major crustal rocks, global fluxes derived from major world rivers, and the relative contribution of the four broad rock types to Earth's exposed land surface (E. *s1*), we estimate a total global denudation rate by abiotic chemistry alone of about 6.1 Tmoles major cations yr$^{-1}$ (*SI 2.5*). An aerobic, microbial-dominated world, i.e., Early Proterozoic (~2.45 Ga), present-day mountain tops, receding glaciers, fresh lava fields and deep biosphere would contribute 11.5% over abiotic (6.8 Tmoles yr$^{-1}$ microbes + abiotic) while one with vascular plants (but without mycorrhizae), would yield 97% of that by microbes (6.6 Tmoles yr$^{-1}$ abiotic + microbes + plant) with a clear net retention in plant biomass of 0.17 Tmoles yr$^{-1}$. A more complex world, e.g. Phanerozoic (0.54 Ga to present), where microbes, fungi and vascular plants evolved symbiotic interactions (0.45 Ga), contributes about 6.2 Tmoles yr$^{-1}$ to the planetary cation cycle, meaning that microbial and fungal components of biosphere accelerate denudation while plants increase retention within evolving ecosystems. Because mountain topography (covering < 10 % of the modern planetary surface) contributes > 50 % of the weathered solutes to oceans [43], our study is most likely to model those regions of primary ecosystem succession and more active soil genesis, as well as the simpler ecosystems and active bedrock surfacing characteristic of earlier periods of Earth history.

## Conclusions

An incipient ecosystem developed sustainably on granular rock substrates using unrecycled nutrients from primary minerals. After plant germination, there was an initial high flux of elements readily available from open mineral structures which lasted 3-4 months and coincided with a dramatic biomass buildup. This was followed by a lower, but steady release of dissolved solutes and an increased plant deficiency in the N:P:K:Mg series over time, which shows the critical connection between an emerging near-surface geosphere and the developing biosphere. Such initial high denudation period supports the notion that in areas of mountain uplift, fresh volcanic fields, or uncovered glacial bedrock with primary ecosystem succession, early mineral exposure, fast bedrock fracturing, and biologically - aided carbonation drawing C from the atmospheric reservoir are contributing larger amounts of nutrients to environment than older, lowland areas. It is also implicit that periods of active orogeny and exposed land early in Earth history have contributed large nutrient fractions to the environment with effects rippling throughout the developing biosphere.

As ecosystem complexity (and development) increased, the substrates retention capacity for water, cations and anions also increased, which had a major impact on the allocation of dissolved ions to exchangeable fraction as well as the Al-silicates and Fe-oxyhydroxides. This is direct evidence of biologically driven incipient secondary mineral nucleation, implying an important effect of an emerging biosphere on nutrient cycles terrestrialization.

The formation of biosignatures was substrate-specific. Inter-elemental patterns emerged under different biota, and varied between weathering pools; nonetheless some element-specific trends were present. The clearest signatures of microbial colonization (compared to abiotic) were in the dissolved and exchangeable pools with the Ca signature in water being consistent across rock types. Compared to microbes alone, vascular plants altered Si stoichiometry across rocks in water, and showed a strong Ti signature in exchangeable and amorphous pools from schist, overlapping variation in other elements. Where present, fungal influence resulted in a Ti signal (different than that of plant) in water across rocks and Ca-K-Al in the exchangeable pool.

Overall, our results aim to more accurately constrain regolith and planetary evolution models where mass contributions from abiotic and different biosphere components are to be precisely known, and to better constrain biosignatures formation on Earth and beyond. Since basalt used in this study could represent Lunar and Mars soil replicates, our results could aid in better understanding the nutrient compartmentalization during terraforming in long-duration planetary missions.

## Methodology

**Rock substrate.** Rocks used in the experiment were collected from Santa Catalina Mountains, Tucson, Arizona (granite and schist), Jemez River Basin, New Mexico (rhyolite) and Merriam crater, Flagstaff, Arizona (cinder basalt). These sites are associated to Critical Zone Observatory and Landscape Evolution Observatory, two large-scale multidisciplinary studies (catchment and hill slope scales) at the University of Arizona. Except for basalt, which was ground at the mining site, the rocks had weathered surfaces removed by a tungsten carbide - tip air hammer, before being crushed in a jaw crusher. Resulted material was dry sieved, then wet sieved using a FRITSCH Vibratory Sieve Shaker (Idar-Oberstein, Germany) to retain the 250-500 µm fraction, passed on a Wilfley gravity water table to remove potential contamination from grinding, mixed and rinsed several times with nanopure-grade water, and dried in air flow ovens at 70°C.

      To identify mineral composition of each rock and map element distribution in minerals, about one gram of granular rock was mounted in an epoxy block, polished to obtain a mirror-like surface, and analyzed for major element distribution and abundances using CAMECA SX100 Ultra and CAMECA SX50 electron probe microanalyzers (Lunar and Planetary Sciences Laboratory, University of Arizona, USA). Minerals were identified using Energy Dispersive Spectrometry (EDS) which simultaneously collects all x-ray wavelengths (energies) emitted by the mineral during the point analysis mode, and their chemical formula inferred from oxygen-normalized element abundances.

      We estimated rock total element contents by total digestion using lithium metaborate/tetraborate fusion - ICP/MS at Activation Laboratories Inc., Ancaster, Ontario, Canada.

**Microbial inoculation.** To avoid introducing significant amount of secondary minerals, a native microbial consortium was extracted from fresh granular basalt collected from a pristine site at Merriam crater,

Flagstaff, N. Arizona (35°20'3.23"N; 111°16'45.48"W). The inoculum was prepared by mixing 1.0 g basalt with 95 mL of sterile ultrapure water[20]. The mixture was vortexed for 2 min to separate rock particles and bath sonicated (VWR Aquasonic 250D model; 120V, 4A, 40Hz) for 2 min to further separate the microbiota. Heterotrophic bacteria abundance was assessed by plating the inoculum on R2A agar. About 90 ml of living inoculum (containing $1.43 \times 10^5$ colony forming units mL$^{-1}$) was passed through 25µm sieve to remove potential native mycorrhiza, and mixed with sand substrate in each treatment replicate except control. In control the microbial extract was sterilized to keep a chemical inoculation similar among treatments.

**Plant and mycorrhizal material.** A low nutrient - tolerant perennial grass, buffalo grass (*Bouteloua dactyloides*) and an arbuscular mycorrhiza (AM) fungus, *Rhizophagus irregularis* (formerly *Glomus intraradices*) were used as model mycorrhizal association in this study. Grass seeds (purchased from Western Native Seed, Colorado, U.S.A.) were dehusked, surface sterilized in 95% ethanol and 2% sodium hypochlorite, rinsed with 0.1 % sodium thiosulfate solution and nanopure water, and pregerminated in sterile water before planting them into granular rock substrates (Fig. s1B, C). Sterile spores of AM (MYKE, Premier Tech Biotechnologies, Canada) were used to inoculate buffalo grass seedlings directly on substrate (Fig. s1 D). Colonization percentage of roots for the experiment period was 0-85%.

**Experiment design and growth conditions.** A modular design comprising six mezocosm chambers, provided with purified air flow (Fig. s1 I) and purified watering systems (Fig. s1H) were set up in the Desert Biome at University of Arizona Biosphere 2 facility in Oracle, Arizona (Fig. s1A). Each chamber contained Plexiglas experimental columns (solid-state bioreactors) 30cm/5cm internal diameter (Fig. s1E) filled with granular rock material, in which 4 biotic treatments were placed: un-inoculated control (C), rock microbes (B), microbes-grass (BG) and grass-microbes-mycorrhiza (BGM). To avoid potential cross-contamination, the control was placed first in the direction of air flow, with the rest of the treatments distributed randomly in the module. Potential experimental edge effect was addressed by using tilted chamber walls.

The experiment was run for 570 days under natural photoperiod, with mean temperature in the biome for the whole period of 19±4°C and relative humidity of 48±19%, at natural $O_2/CO_2$ saturation conditions. Sterile nanopure (18 MΩ) water was applied biweekly to each column by 120 mL syringe (Fig. s1H) at a rate of 4 mL s$^{-1}$ (a total of 100-120 mL column$^{-1}$ added each time) by a dripping system designed to avoid preferential flow (Fig. s1G), bringing granular rock profiles to half-full capacity each time. The syringe was pre-sterilized for each watering and the connection sterilized using ethanol. Pore water was sampled gravimetrically (Fig. s1F) biweekly for the first two months, and monthly thereafter, generating 30-50mL column$^{-1}$.

**Plant elemental analysis.** After 135, 253, 465 and 583 days, triplicate columns were sacrificed, and plants were harvested and separated into above and below-ground biomass. Roots were gently washed in 18 MΩ nanopure water three times to remove rock grains attached to their surface. A further microscope examination of the roots confirmed particle removal. Dry biomass was determined for shoot and root after oven-drying at 70°C for 72h. About 30 randomly selected root fragments (1cm) were stained using trypan blue to determine mycorrhizal infection rate (by optical microscopy) in each planted column[20]. Shoot and root were digested separately in 70% $HNO_3$ Aristar plus-BDH and 40% $H_2O_2$ J.T. Baker's Ultrex (1:1 mixture, microwave assisted), and analyzed for Na, Mg, Al, Si, P, K, Ca, Ti, Mn, and Fe concentrations by inductively coupled plasma mass spectrometry (ICP-MS; Perkin Elmer, Elan DRC-II). Quality control checks included sample blanks, certified reference material (apple leaves CRM 1515) and triplicate sample analysis.

**Analysis of drainage solution.** The volume of drainage solution was measured each water sampling events and water consumption capacity (evaporated and uptaken by biota) was estimated as the difference between input and output volumes. Samples were further analyzed for dissolved inorganic and non-purgeable organic carbon (IC and NPOC), and total nitrogen (TN) by Shimadzu TOC-L system, and acidified and analyzed for same suite of elements as in plants by ICP-MS. Solution pH and conductivity were also measured each sampling time. Major anions, (bromide, sulphate, phosphate, fluoride and chloride) were analyzed by ion chromatography (Dionex). Additionally, pH and conductivity were measured by electronic pH and conductivity probes for each sample.

**Secondary solid phase analysis (sequential extraction).** At the end of the 20 months experiment, bulk rock samples (3 replicates per each rock and each treatment) were subjected to a two-step operationally-defined chemical extraction of (step 1) easily bioavailable/exchangeable and carbonates (by 0.2 M ammonium acetate adjusted to pH 4.5), and (step 2) amorphous and Fe(III) oxyhydroxides (using 0.2 M ammonium oxalate adjusted to pH 3.0) following ref.[44].

One gram of homogenized granular rock was added to 50 ml polypropylene centrifuge tubes. To extract the exchangeable fraction, 40 ml of 0.2 M ammonium acetate (NH4OAc) was added to sample, vortexed to assure suspension and shaken (7 rpm) at room temperature on a reciprocal shaker (VWR Advanced 3750 shaker) for 60 min. Samples were further centrifuged at 4700 rpm for 20 min to separate supernatant and pellet. The supernatant was filtered through PALL GHP Acrodisc 25 mm 0.45 µm syringe filters (prewashed with nitric acid) into VWR metal-free tubes. The pellet was rinsed with 15 ml deionized water, vortex to resuspend and centrifuged for 20 min. The new supernatant was filtered and saved in VWR tubes pre-washed with 1% nitric acid solution. Mass of each tube and pellet were recorded before step 2 (i.e., amorphous fraction extraction).

The amorphous-to-poorly crystalline fraction (step 2) was extracted by adding 40 ml of 0.2 M ammonium oxalate adjusted to pH 3.0 to pellet resulted from step 1, vortexed to resuspend, shaken at room temperature for 2 hours and 7 rpm and further centrifuged for 20 min at 4700 rpm. The supernatant was filtered through 0.45µm PALL GHP Acrodisc syringe filters prewashed with nitric acid. Filtrates were captured in preweighted VWR tube. Pellets were rinsed with 15 ml 0.1 M acetic acid, vortexed, centrifuged at 4700 rpms for 20 min and collected in clean VWR tubes.

Extracted solutions (supernatant) from each step were acidified with 50% Omni-trace nitric acid to pH 2 and analyzed by ICP-MS for same suite of elements as in plants.

**Modeling of precipitation condition.** To evaluate conditions for element precipitation in pore waters, an analysis of ionic equilibrium in the aqueous pore environment was applied to the average element concentrations (across biological treatments) normalized to water balance (input – output), for each rock, using Visual Minteq v. 3.1 (https://vminteq.lwr.kth.se/). The model computes the % distribution among dissolved and adsorbed ion species for each element, and derives ion saturation indexes with respect to possible precipitated phases.

**Weathering budget calculation and statistical analysis.** Nutrient budget in mass balance analysis was determined by summing the total element masses in the different weathering pools: drainage solution, plant uptake (root and shoot), and solid phases (exchangeable and poorly crystalline). Results were expressed in moles throughout the analyses, to allow element abundance comparisons.

Preferential leaching/denudation was calculated by dividing total mass element in water to total mass element in original rock and it is assumed to be a reflection of weathering capacity of different rock minerals. Preferential uptake was given by dividing total element in plant (mass) to total removal/leaching in water (mass).

After eliminating outliers (discussion on statistical behavior of a number of columns is in SI 1.2) significance (*; p<0.05) of treatment effect on preferential element denudation was derived stepwise (C-B, B-BG, and BG-BGM) from nonparametric Mann-Whitney U test comparisons with Dunn-Sidak adjustment (which corrects the significance for multiple comparison error). To evaluate biotic and rock effects on preferential uptake as well as their effect on different weathering pools ANOVA with Fisher's least significant difference (LSD) posthoc test were applied on inter-treatment comparisons. Principal component analysis (PCA) together with Automatic Linear (regression) Model (ALM) helped to find important correlates in pore water element contents, which could explain their dissolution behavior. PCA is a multivariate technique that clusters measured variables into functional groups denoting independent processes, and has successfully been used in various studies on element geochemical sources[22,45]. ALM is an improved multiple regression model, which provides an automatic data preparation algorithm for big data sets, including removing outliers (3σ from the mean cutoff), and identifying the most influential predictor sets[46]. In ALM, regression factor scores from PCA were modeled against predictors displaying significant (*p*<0.05) Pearson correlation with PCA variables. To maximize the model accuracy, a bootstrap resampling method was used.

Statistical tests were conducted in SPSS 17.0 following the assumption of a general linear model factorial design. Mass balance differences among rocks and treatments were analyzed with one way ANOVA and LSD multiple comparisons testing for effects of rock type and AM/non-AM treatment type.

## References

1. Kaspari, M. & Powers, J. S. Biogeochemistry and Geographical Ecology: Embracing All Twenty-Five Elements Required to Build Organisms*. *Am. Nat.* **188,** S000–S000 (2016).
2. Zaharescu, D. G. *et al.* Riparian vegetation in the alpine connectome: Terrestrial-aquatic and terrestrial-terrestrial interactions. *Sci. Total Environ.* **601–602,** 247–259 (2017).
3. Vernadsky, V. *Biosfera*. (Nauch, 1926).
4. Richter, D. D. & Billings, S. A. One physical system: Tansley's ecosystem as Earth's critical zone, Tansley review. *New Phytol.* **206,** 900–912 (2015).
5. Zaharescu, D. G., Hooda, P. S., Burghelea, C. I. & Palanca-Soler, A. A Multiscale Framework for Deconstructing the Ecosystem Physical Template of High-Altitude Lakes. *Ecosystems* **19,** 1064–

1079 (2016).

6.    Hazen, R. M. *et al.* Mineral evolution. *Am. Mineral.* **93,** 1693–1720 (2008).

7.    Bennett, P. C., Rogers, J. R., Choi, W. J. & Hiebert, F. K. Silicates, Silicate Weathering, and Microbial Ecology. *Geomicrobiol. J.* **18,** 3–19 (2001).

8.    Liermann, L. J., Kalinowski, B. E., Brantley, S. L. & Ferry, J. G. Role of bacterial siderophores in dissolution of hornblende. *Geochim. Cosmochim. Acta* **64,** 587–602 (2000).

9.    Turner, B. L., Lambers, H., Condron, L. M., Cramer, M. D. & Smith, S. E. Soil microbial biomass and the fate of phosphorus during long-term ecosystem development. *Plant Soil* **367,** 225–234 (2013).

10.   Welch, S. A., Barker, W. W. & Banfield, J. F. Microbial extracellular polysaccharides and plagioclase dissolution. *Geochimica et Cosmochimica Acta* 1405–1419 (1999). at <http://ac.els-cdn.com/S0016703799000319/1-s2.0-S0016703799000319-main.pdf?_tid=2c0e309e-7c3a-11e4-8c26-00000aab0f6b&acdnat=1417755181_8a3ce52f6aed25f630114997eb1e0f64>

11.   Adeyemi, A. O. & Gadd, G. M. Fungal degradation of calcium-, lead- and silicon-bearing minerals. *Biometals* 269–281 (2005). at <http://download.springer.com/static/pdf/276/art%253A10.1007%252Fs10534-005-1539-2.pdf?auth66=1417809975_39a11ef32803c7da72e2927e7a479a04&ext=.pdf>

12.   Gadd, G. M. Presidential address Geomycology : biogeochemical transformations of rocks , minerals , metals and radionuclides by fungi , bioweathering and bioremediation. **111,** 3–49 (2007).

13.   Sterflinger, K. Fungi as Geologic Agents. *Geomicrobiol. J.* **17,** 97–124 (2000).

14.   Lambers, H., Mougel, C., Jaillard, B. & Hinsinger, P. Plant-microbe-soil interactions in the rhizosphere : an evolutionary perspective. *Plant Soil* (2009). doi:10.1007/s11104-009-0042-x

15.   Hawes, M. C., Brigham, L. A., Wen, F., Woo, H. H. & Zhu, Y. Function of root border cells in plant health: Pioneers in the rhizosphere. *Annu. Rev. Phytopathol* **36,** 311–27 (1998).

16.   Hinsinger, P. How do plant roots acquire mineral nutrients ? Chemical processes in volved in the rhizosphere. *Adv. Agron.* **64,** 225–265 (1998).

17.   Terrer, C. *et al.* Mycorrhizal association as a primary control of the $CO_2$ fertilization effect. *Science* **353,** 72–4 (2016).

18.   Storkey, J. *et al.* Grassland biodiversity bounces back from long-term nitrogen addition. *Nature* **528,** 401–404 (2015).

19.   Dixon, J. L. & von Blanckenburg, F. Soils as pacemakers and limiters of global silicate weathering. *Comptes Rendus Geosci.* **344,** 597–609 (2012).

20.   Heimsath, A. M., DiBiase, R. A. & Whipple, K. X. Soil production limits and the transition to bedrock-dominated landscapes. *Nat. Geosci.* **5,** 210–214 (2012).

21.   Burghelea, C. I. *et al.* Mineral nutrient mobilization by plant from rock: Influence of rock type and arbuscular mycorrhiza. *Biogeochemistry* **124,** 187–203 (2015).

22.   Zaharescu, D. G. *et al.* Ecosystem Composition Controls the Fate of Rare Earth Elements during Incipient Soil Genesis. *Sci. Rep.* **7,** 1–15 (2017).

23.   Brantley, S. L. *et al.* Twelve testable hypotheses on the geobiology of weathering. *Geobiology* **9,** 140–65 (2011).

24.   Balogh-Brunstad, Z. *et al.* Chemical weathering and chemical denudation dynamics through ecosystem development and disturbance. *Global Biogeochem. Cycles* **22,** 1–11 (2008).

25.   Gadd, G. M. in *Global Ecology* 1709–1717 (2008).

26.   Hausrath, E. M. *et al.* Short- and long-term olivine weathering in Svalbard: implications for Mars. *Astrobiology* **8,** 1079–92 (2008).

27.   Levenson, Y. & Emmanuel, S. Repulsion between calcite crystals and grain detachment during

water–rock interaction. *Geochemical Perspect. Lett.* **3,** 133–141 (2017).

28. Des Marais, D. J. Carbon, nitrogen and sulfur in Apollo 15, 16 and 17 rocks. in *Proc. 9th Lunar Planet. Sci. Conf.* 2451–2467 (1978).

29. Holloway, J. M. & Dahlgren, R. A. Nitrogen in rock: Occurrences and biogeochemical implications. *Global Biogeochem. Cycles* **16,** 65 1-17 (2002).

30. Sakai, H., Marais, D. J. Des, Ueda, A. & Moore, J. G. Concentrations and isotope ratios of carbon, nitrogen and sulfur in ocean-floor basalts. *Geochim. Cosmochim. Acta* **48,** 2433–2441 (1984).

31. Quirk, J. *et al.* Evolution of trees and mycorrhizal fungi intensifies silicate mineral weathering. *Biol. Lett.* **8,** 1006–1011 (2012).

32. Hartmann, J. Bicarbonate-fluxes and CO2-consumption by chemical weathering on the Japanese Archipelago - Application of a multi-lithological model framework. *Chem. Geol.* **265,** 237–271 (2009).

33. Hermans, C. *et al.* Systems analysis of the responses to long-term magnesium deficiency and restoration in Arabidopsis thaliana. *New Phytol.* **187,** 132–44 (2010).

34. Sterner, R. W. & Elser, J. J. *Ecological Stoichiometry: The Biology of Elements from Molecules to the Biosphere.* (Princeton University Press, 2002).

35. Lorimer, G. H., Badger, M. R. & Andrews, T. J. The activation of ribulose-1,5-bisphosphate carboxylase by carbon dioxide and magnesium ions. Equilibria, kinetics, a suggested mechanism, and physiological implications. *Biochemistry* **15,** 529–536 (1976).

36. Kai, Y., Matsumura, H. & Izui, K. Phosphoenolpyruvate carboxylase: Three-dimensional structure and molecular mechanisms. *Arch. Biochem. Biophys.* **414,** 170–179 (2003).

37. Leigh, R. A. & Jones, R. G. W. A Hypothesis Relating Critical Potassium Concentrations for Growth to the Distribution and Functions of this Ion in the Plant Cell. *New Phytol.* **97,** 1–13 (1984).

38. Hopkins, W. G. & Huner, N. P. A. *Introduction to Plant Physiology. Agronomy Journal* **43,** (John Wiley & Sons, 2009).

39. Buss, H., Brantley, S. & Liermann, L. Nondestructive Methods for Removal of Bacteria from Silicate Surfaces. *Geomicrobiol. J.* **20,** 25–42 (2003).

40. Tazaki, K. *Clays , microorganisms, and biomineralization. Handbook of Clay Science* **1,** (2006).

41. Estes, E. R., Andeer, P. F., Nordlund, D., Wankel, S. D. & Hansel, C. M. Biogenic manganese oxides as reservoirs of organic carbon and proteins in terrestrial and marine environments. *Geobiology* **15,** 158–172 (2017).

42. Lenton, T. M., Crouch, M., Johnson, M., Pires, N. & Dolan, L. First plants cooled the Ordovician. *Nat. Geosci.* **5,** 86–89 (2012).

43. Larsen, I. J., Montgomery, D. R. & Greenberg, H. M. The contribution of mountains to global denudation. *Geology* **42,** 527–530 (2014).

44. Dold, B. Speciation of the most soluble phases in a sequential extraction procedure adapted for geochemical studies of copper sulfide mine waste. *J. Geochemical Explor.* **80,** 55–68 (2003).

45. Zaharescu, D. G., Hooda, P. S., Fernandez, J., Soler, A. P. & Burghelea, C. I. On the arsenic source mobilisation and its natural enrichment in the sediments of a high mountain cirque in the Pyrenees. *J. Environ. Monit.* **11,** 1973–81 (2009).

46. Yang, H. The Case for Being Automatic: Introducing the Automatic Linear Modeling (LINEAR) Procedure in SPSS Statistics. *Mult. Linear Regres. Viewpoints* **39,** 27–37 (2013).

## Aknowledgements

This research was funded by National Science Foundation (NSF) grant EAR-1023215 "ETBC: Plant-microbe-mineral interaction as a driver for rock weathering and chemical denudation". Additional support came from NSF EAR-0724958 and EAR-1331408 grants that support the Catalina - Jemez Critical Zone

Observatory (CZO), the Biosphere 2 REU program, NSF EAR-1263251 and NSF EAR-1004353 (http://www.b2science.org/outreach/reu), United States-Mexico Commission for Educational and Cultural Exchange (COMEXUS): the Fulbright-Garcia Robles Scholarship program, Thomas R. Brown Foundation endowment to University of Arizona, We also acknowledge the support from NSF EAR-1411609 "ELT: Collaborative Research: Beyond the boring billion: Late Proterozoic glaciation, oxygenation and the proliferation of complex life" and NASA Astrobiology Institute "CAN7: Alternative Earths. Explaining Persistent Inhabitation on a Dynamic Early Earth".

Use of the Stanford Synchrotron Radiation Lightsource, SLAC National Accelerator Laboratory, is supported by the U.S. Department of Energy, Office of Science, Office of Basic Energy Sciences under Contract No. DE-AC02-76SF00515.

We are deeply thankful to John Adams, Julia Perdrial, Nicolas Perdrial, Travis Huxman, Nate Abramson, Viktor Polyakov, Yadi Wang, Elizabeth K. Nadeau, Juliana Gil Loiaza, Jake Kelly, Vanessa Yubeta, Lauren Guthridge, Mathew Clark, James Olmid, Guillermo Molano, Andrew Toriello, Nicolas Sertillanges, Arturo Jacobo, Julie Neilson, Kenneth Kanipe, Rebecca Lybrand, Ariel Lorenzo, Kalee Vasquez, Ashi Bhaat, Tekatrianna Schulte-Evans, and the multiple other collaborators for their valuable logistics, field, lab and theoretical contributions.

## Authors credit

1. Grant preparation: KD, JC, RMM
2. Study design and writing (text, figures, tables and SI): DGZ, KD, JC, CTR, CIB
3. Material preparation and initial tests: DGZ, CIB, JKP, EAH
4. Mezocosm setup and running: DGZ, JKP, CIB
5. Sample collection and preparation: DGZ, CIB, JKP, SS, EG, MG, EM
6. Analytics of liquid, solid and biological materials: DGZ, CIB, JKP, KJD, EAH, MKA, SS, EEG, MOVI, MAPM, RCM, ERN
7. Data analysis: DGZ, CIB, JKP, KJD, SS, EM, EEG, MOVI, MAPM, KL
   Review: DGZ, KD, JC, CTR, CIB, ERN

## Additional Information

**Competing interests**. The authors declare no competing interests.

# SUPPLEMENTARY INFORMATION

## SI 1 Supplementary Methods

### SI 1.1 Experimental setup

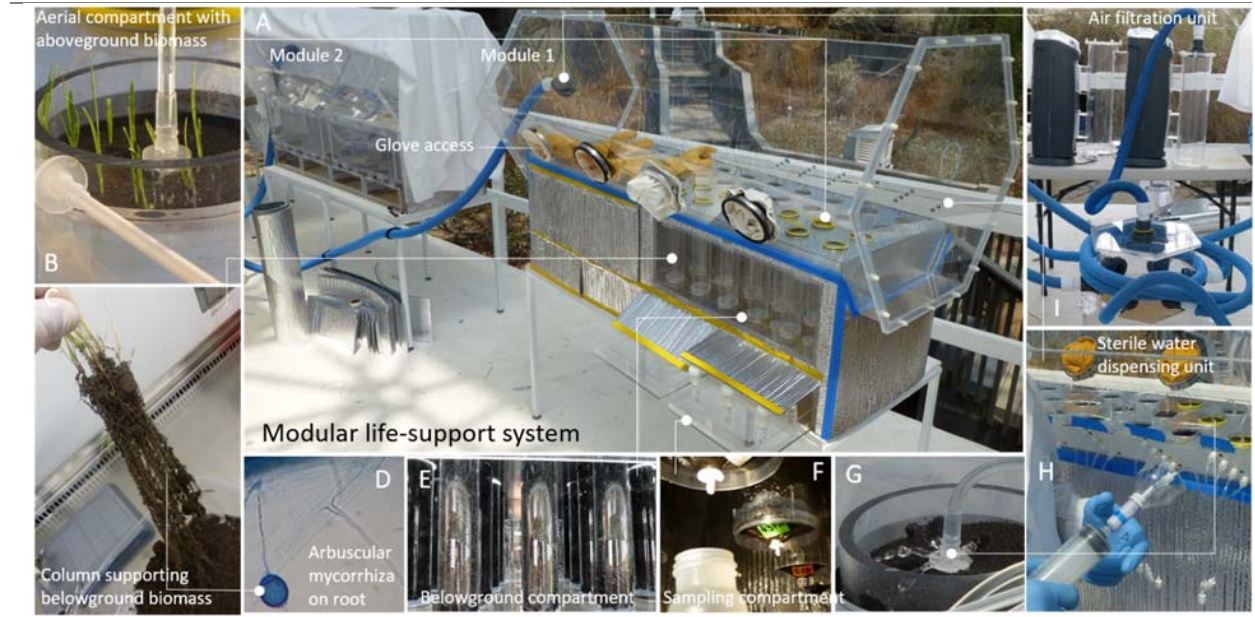

*Figure s1* **The experimental setup.** Mezocosm chambers represented by six growing units - rockubators (A) in the Desert Biome of Biosphere-2, Arizona. Granular rock - filled columns, 30 x 5 cm long, extend from above - ground (B; Buffalo grass shoots in basalt), to below - ground (C – E; Buffalo grass roots in rhyolite), and to pore water sampling (F) compartments. Purified water and air are delivered through sterile syringe (G and H), and air-purification system (I).

**SI 1.2 Behavior of treatment replicate.** In experimental biogeosciences replicate treatment architecture and associated comparison statistics are required to rule out bias in quantitative results and interpretation, and give predictability power to hypotheses. Treatment triplicates are used as a rule of thumb to establish facts and hence draw conclusions on biogeochemical processes. In systems that are highly sensitive to small variations in processes involved, i.e. high noise, such as our study, the rule-of-three is harder to achieve, hence the reliability of analyzed parameters can be lower. Results from our time series plotting of pore water element content showed that over the studied time a limited number of columns behaved independently of the rest in their triplicate treatment behavior. While they represent real processes, the cause of these unique behaviors has not been identified. To preserve statistical power across the whole experiment, these columns were removed from the general statistical analysis.

## SI 2 Supplementary Results

### SI 2.1 Substrate

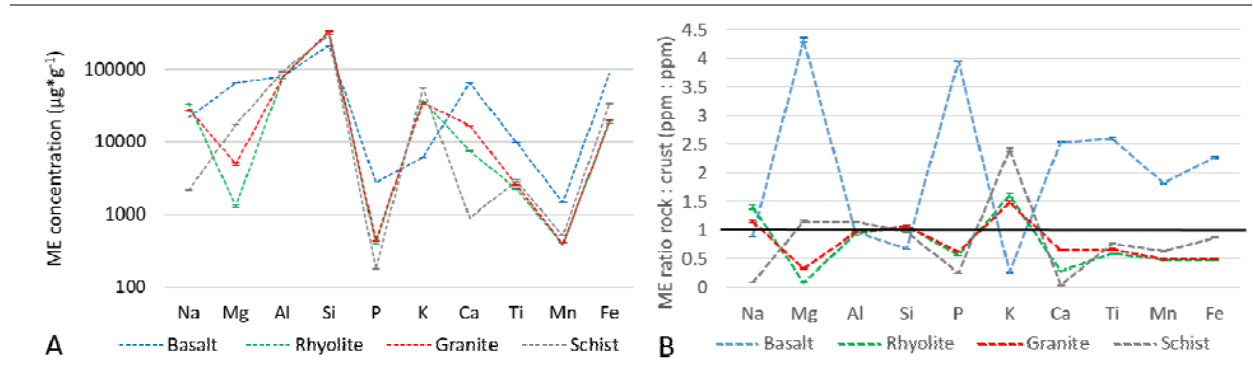

*Figure s2* **Substrates elemental composition.** Variability of major element (ME) concentrations in initial rocks (A; log scale), and their values relative to upper continental crust averages[‡], with a reference black line set at unity (B).

### SI 2.2 Pore water

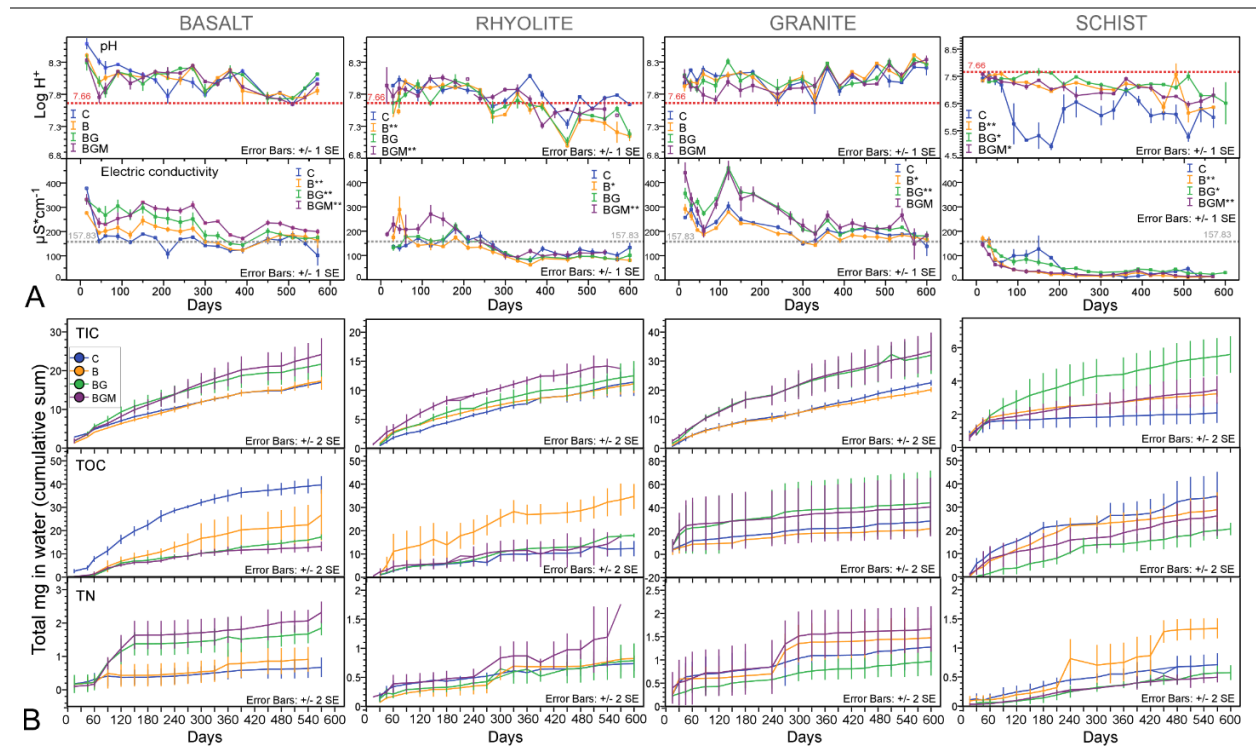

*Figure s3* **Changes in water physico - chemical variates**. Time lapse of (A) pH and electrical conductivity (horizontal dash line is set at the average value for all rocks and treatments), and (B) total anions in pore waters across biotic treatments in each rock. Significance of overall treatment effect in A (legend), was determined at *1SE and **2SE in increasing order of treatment complexity for C-B, B-BG and BG-BGM. A. Biological treatment C, control; B, microbes; BG, microbes-grass; BGM, microbes-grass-mycorrhiza.

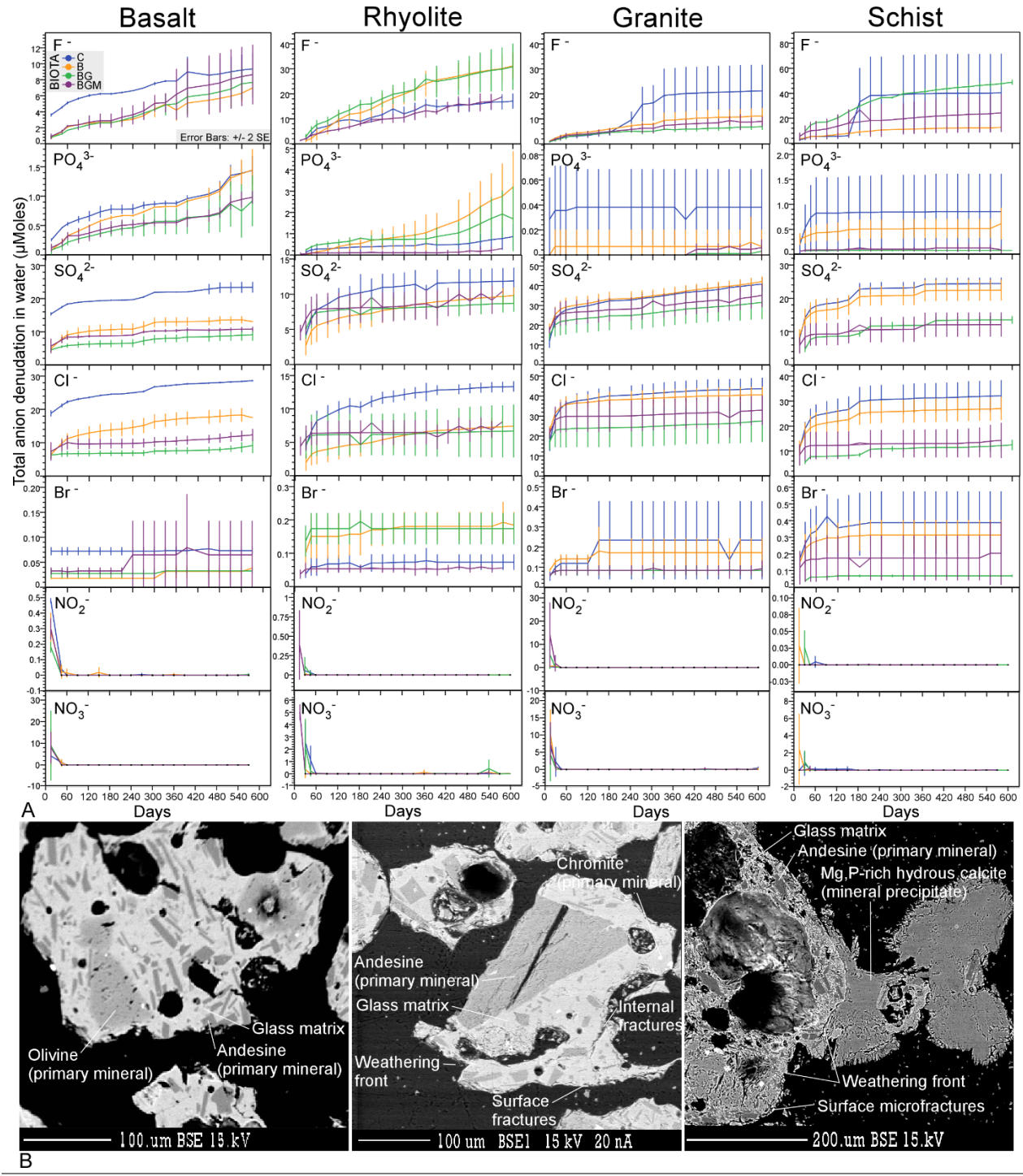

**Figure s4** **Anion leaching in time and the physical weathering effect.** (A) Cumulative sum (except nitrite and nitrate) of dissolved anions in pore waters over time, across biotic treatments in each substrate. Legend symbols, for all plots: C, control; B, microbes; BG, microbes-grass; BGM, microbes-grass-mycorrhiza. (B) Electrode microprobe images of basalt grain weathering front at time 0 (left) and end of experiment (center, abiotic control), compared to naturally weathered material from Merriam crater, Arizona (right, with representation of microfractures and precipitated secondary minerals).

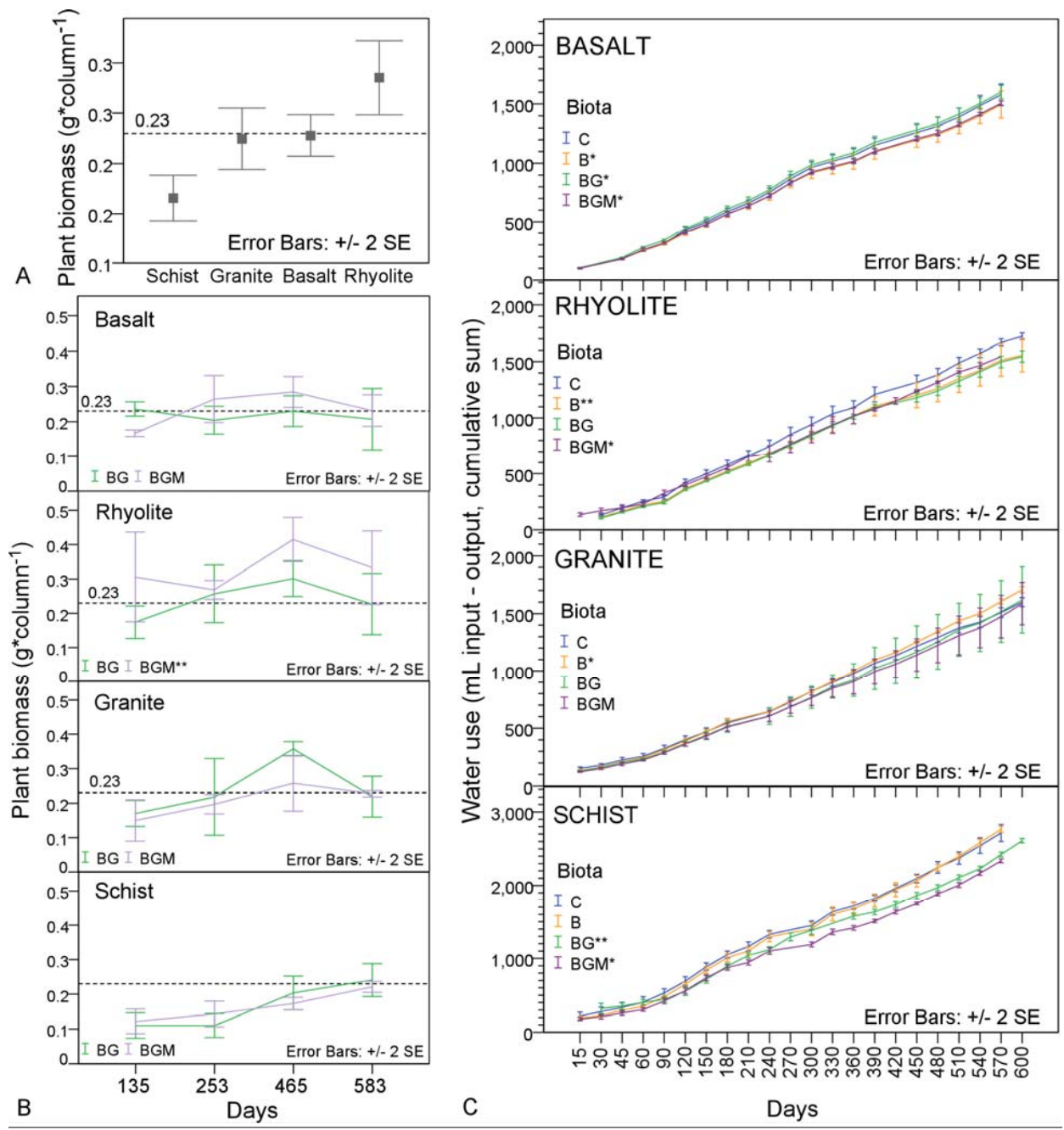

*Figure s5* **Changes in plant growth and ecosystem water consumption.** Substrate effect on plant biomass across all four sample events (A) together with time change in plant biomass (B), and total column water consumption (C)- estimated as volume of irrigation input – volume of extracted pore water, over the course of the experiment. Significance of treatment effect displayed in legends of subfigures B and C was determined at *1SE and **2SE, stepwise for C-B, B-BG and BG-BGM. C, control; B, microbes; BG, microbes-grass; BGM, microbes-grass-mycorrhiza.

*Table s1* **Major pore water physico-chemical parameters.** Descriptives of water consumption, pH, measures of C, N and major ion contents in water leached from four silicate rocks during 20 months biological weathering, across biotic treatments. Except for pH and EC, which show column averages, values are sums per column for the 60 sampling events, and represent total denudation at the end of the experiment. Means and standard errors (SE) were derived from column triplicates. C, control; B, microbial; BG, microbial–grass; and BGM, microbial–grass–arbuscular mycorrhiza.

| Variable | Treatment | Basalt | | | Rhyolite | | | Granite | | | Schist | | |
|---|---|---|---|---|---|---|---|---|---|---|---|---|---|
| | | Mean | ± | SE | Mean | ± | SE | Mean | ± | SE | Mean | ± | SE |
| Water balance (input-output; ml) | C | 1579.23 | ± | 45.27 | 1763.58 | ± | 14.25 | 1639.50 | ± | 19.13 | 2754.89 | ± | 57.46 |
| | B | 1515.66 | ± | 38.88 | 1595.81 | ± | 73.20 | 1748.13 | ± | 13.13 | 2803.11 | ± | 21.52 |
| | BG | 1601.41 | ± | 28.35 | 1584.14 | ± | 25.17 | 1658.15 | ± | 142.89 | 2650.10 | ± | 14.65 |
| | BGM | 1506.08 | ± | 10.29 | 1512.08 | ± | 41.03 | 1627.17 | ± | 91.89 | 2377.25 | ± | 13.05 |
| pH | C | 8.03 | ± | 0.013 | 7.80 | ± | 0.026 | 8.04 | ± | 0.018 | 6.27 | ± | 0.23 |
| | B | 8.01 | ± | 0.0072 | 7.61 | ± | 0.027 | 8.07 | ± | 0.011 | 7.06 | ± | 0.054 |
| | BG | 8.01 | ± | 0.030 | 7.64 | ± | 0.043 | 8.06 | ± | 0.026 | 7.24 | ± | 0.049 |
| | BGM | 7.99 | ± | 0.020 | 7.77 | ± | 0.035 | 7.99 | ± | 0.037 | 7.06 | ± | 0.079 |
| EC (µs/cm) | C | 167.09 | ± | 3.16 | 134.30 | ± | 6.56 | 211.59 | ± | 4.85 | 59.23 | ± | 5.75 |
| | B | 190.49 | ± | 12.89 | 123.89 | ± | 10.38 | 198.93 | ± | 3.20 | 41.92 | ± | 2.14 |
| | BG | 228.27 | ± | 11.14 | 127.33 | ± | 8.33 | 265.90 | ± | 1.00 | 56.48 | ± | 1.80 |
| | BGM | 248.05 | ± | 2.52 | 158.32 | ± | 12.01 | 267.70 | ± | 20.43 | 39.75 | ± | 4.81 |
| TC (mg) | C | 56.65 | ± | 1.86 | 23.72 | ± | 2.28 | 51.14 | ± | 4.26 | 36.77 | ± | 5.00 |
| | B | 40.95 | ± | 4.00 | 45.85 | ± | 3.45 | 42.21 | ± | 1.88 | 32.07 | ± | 3.70 |
| | BG | 38.93 | ± | 1.47 | 30.45 | ± | 0.84 | 76.34 | ± | 16.95 | 26.19 | ± | 1.76 |
| | BGM | 37.43 | ± | 1.11 | 28.53 | ± | 2.03 | 73.92 | ± | 15.90 | 29.65 | ± | 3.88 |
| TN (mg) | C | 0.677 | ± | 0.15 | 0.738 | ± | 0.066 | 1.275 | ± | 0.23 | 0.716 | ± | 0.10 |
| | B | 0.914 | ± | 0.19 | 0.825 | ± | 0.026 | 1.479 | ± | 0.23 | 1.339 | ± | 0.086 |
| | BG | 1.741 | ± | 0.13 | 0.788 | ± | 0.15 | 0.969 | ± | 0.15 | 0.576 | ± | 0.064 |
| | BGM | 2.323 | ± | 0.16 | 1.211 | ± | 0.28 | 1.667 | ± | 0.25 | 0.498 | ± | 0.045 |
| Anions (µm) | C | 62.96 | ± | 1.20 | 43.27 | ± | 0.88 | 105.94 | ± | 2.26 | 98.17 | ± | 20.05 |
| | B | 40.01 | ± | 0.95 | 51.84 | ± | 4.14 | 94.17 | ± | 1.71 | 63.03 | ± | 3.63 |
| | BG | 26.91 | ± | 1.45 | 48.48 | ± | 2.87 | 66.20 | ± | 9.90 | 75.05 | ± | 0.73 |
| | BGM | 32.84 | ± | 2.46 | 32.45 | ± | 2.92 | 77.15 | ± | 6.67 | 51.00 | ± | 6.42 |
| Cations (µm) | C | 1436.90 | ± | 10.75 | 983.48 | ± | 80.50 | 1485.22 | ± | 35.28 | 486.16 | ± | 27.71 |
| | B | 1313.17 | ± | 34.02 | 1168.93 | ± | 15.18 | 1256.16 | ± | 21.87 | 363.38 | ± | 11.99 |
| | BG | 1438.39 | ± | 79.19 | 1221.83 | ± | 70.35 | 1631.96 | ± | 109.61 | 623.26 | ± | 35.67 |
| | BGM | 1586.86 | ± | 158.50 | 1012.49 | ± | 16.57 | 1707.76 | ± | 149.20 | 407.73 | ± | 29.49 |

*Table s2.* **Drivers of abiotic dissolution.** Relationship between dissolved elements and principal components (PC) of Principal Component Analysis (PCA; in brackets), together with the predictor variables of PCs as derived from Automatic Linear Model (ALM). ALM accuracy represents total variance explained by predictors, and model values are significant at $p > 95$ %. PCs are on different shades of grey. Anions and predictors were input into the model as molar concentrations, while cations are rock-normalized values (representing preferential loss, *). Input values were averaged across treatment triplicates. PCA rotation method: Varimax with Kaiser normalization.

| Basalt | PC (loading) | PC Predictor (% importance) | Model accuracy (%) | Rhyolite | PC (loading) | PC Predictor (% importance) | Model accuracy (%) |
|---|---|---|---|---|---|---|---|
| NO$_3$ | 1 (0.99) | | | NO$_2$ | 1 (0.95) | | |
| SO$_4^{2-}$ | 1 (0.99) | | | NO$_3$ | 1 (0.95) | | |
| Cl- | 1 (0.98) | | | Br- | 1 (0.92) | | |
| NO$_2$ | 1 (0.98) | | | *Fe | 1 (0.92) | | |
| Br- | 1 (0.97) | | | SO$_4^{2-}$ | 1 (0.89) | H$_2$CO$_3$ (76), pH (24) | 17.0 |
| *Na | 1 (0.97) | Carbonate (54), pH (27), Bicarbonate (19) | 62.2 | Cl- | 1 (0.89) | | |
| F- | 1 (0.96) | | | PO$_4^{3-}$ | 1 (0.83) | | |
| *Al | 1 (0.95) | | | *P | 1 (0.75) | | |
| *Fe | 1 (0.94) | | | *Si | 1 (0.72) | | |
| *K | 1 (0.90) | | | *Al | 1 (0.72) | | |
| *Si | 1 (0.86) | | | *Mg | 2 (0.91) | | |
| *Ti | 1 (0.80) | | | *Ca | 2 (0.91) | pH (72), Carbonate (28) | 12.0 |
| *P | 1 (0.76) | | | *K | 2 (0.81) | | |
| PO$_4^{3-}$ | 1 (0.72) | | | *Mn | 3 (-0.83) | | |
| *Mg | 2 (0.94) | | | F- | 3 (0.74) | Carbonate (66), pH (34) | 27.7 |
| *Ca | 2 (0.90) | | | *Na | 3 (0.65) | | |
| *Mn | 3 (0.91) | TOC (-,100) | 44.6 | *Ti | 4 (0.96) | | |

Total variance explained (%): PC1 (70.2), PC2 (12), PC3 (8).

Total variance explained (%): PC1 (46), PC2 (17.6), PC3 (15), PC4 (7.9).

| Granite | PC (loading) | PC Predictor (% importance) | Model accuracy (%) | Schist | PC (loading) | PC Predictor (% importance) | Model accuracy (%) |
|---|---|---|---|---|---|---|---|
| NO$_2$ | 1 (0.97) | | | *Na | 1 (0.99) | | |
| Cl- | 1 (0.96) | | | *P | 1 (0.98) | | |
| SO$_4^{2-}$ | 1 (0.96) | | | Cl- | 1 (0.96) | | |
| NO$_3$ | 1 (0.93) | TOC (100) | 47.40 | Br- | 1 (0.95) | Carbonate (42), Bicarbonate (41), pH (9), TOC (5), H$_2$CO$_3$ (3) | 97.0 |
| PO$_4^{3-}$ | 1 (0.88) | | | SO$_4^{2-}$ | 1 (0.94) | | |
| *Na | 1 (0.83) | | | PO$_4^{3-}$ | 1 (0.93) | | |
| *Si | 1 (0.81) | | | *Si | 1 (0.73) | | |
| *K | 2 (0.94) | | | *Ca | 1 (0.71) | | |
| *Ca | 2 (0.88) | | | NO$_3$ | 1 (0.65) | | |
| *Fe | 2 (0.67) | | | *Mn | 2 (0.93) | | |
| Br- | 2 (-0.51) | | | *Mg | 2 (0.83) | pH (56), TOC (30), H$_2$CO$_3$ (14) | 84.0 |
| *Al | 3 (-0.90) | | | F- | 2 (0.80) | | |
| *Mg | 3 (0.75) | TOC(54), Bicarbonate (46) | 47.70 | *K | 2 (0.78) | | |
| *P | 3 (0.71) | | | *Fe | 3 (0.89) | | |
| *Mn | 3 (-0.60) | | | *Ti | 3 (0.79) | | |
| *Ti | 4 (0.93) | | | *Al | 3 (0.75) | | |
| F- | 5 (0.95) | H$_2$CO$_3$(68), pH(32) | 32.10 | NO$_2$ | 4 (0.90) | | |

Total variance explained (%): PC1 (38), PC2 (19.3), PC3 (15.8), PC4 (8), PC5 (7.8).

Total variance explained (%): PC1 (44.9), PC2 (21.4), PC3 (13.4), PC4 (7.5).

**Table s3. Drivers of biotic dissolution (a).** Grouping of dissolved elements by Principal Component Analysis, their correlation coefficient to the group (PC loading, in brackets), % total variance explained by each PC (table bottom), and relative importance (%) of PC predictors, derived from Automatic Linear Model (ALM). Total variance explained by predictors (%) is significant at $p > 95$ %. Anions and predictors (pH, H$^+$, H$_2$CO$_3$, carbonate, bicarbonate, total organic carbon-TOC) were input into the model as molar concentrations, while cations as rock-normalized values (representing preferential loss, *). Input values were averaged across treatment triplicates. PCA rotation method: Varimax with Kaiser normalization. Treatments: B, microbial; BG, microbial–grass; and BGM, microbial–grass–arbuscular mycorrhiza.

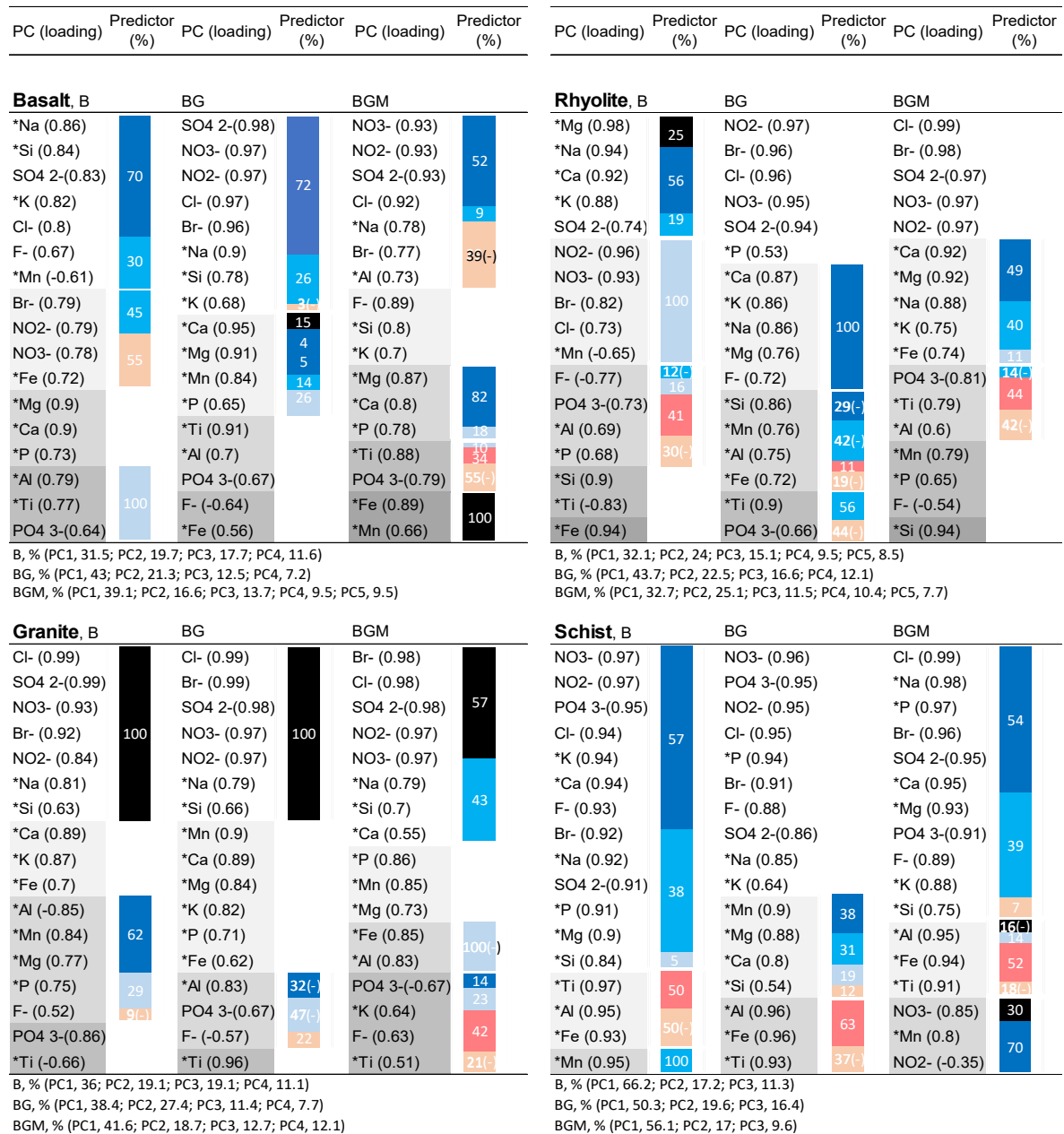

**Basalt**, B

B, % (PC1, 31.5; PC2, 19.7; PC3, 17.7; PC4, 11.6)
BG, % (PC1, 43; PC2, 21.3; PC3, 12.5; PC4, 7.2)
BGM, % (PC1, 39.1; PC2, 16.6; PC3, 13.7; PC4, 9.5; PC5, 9.5)

**Rhyolite**, B

B, % (PC1, 32.1; PC2, 24; PC3, 15.1; PC4, 9.5; PC5, 8.5)
BG, % (PC1, 43.7; PC2, 22.5; PC3, 16.6; PC4, 12.1)
BGM, % (PC1, 32.7; PC2, 25.1; PC3, 11.5; PC4, 10.4; PC5, 7.7)

**Granite**, B

B, % (PC1, 36; PC2, 19.1; PC3, 19.1; PC4, 11.1)
BG, % (PC1, 38.4; PC2, 27.4; PC3, 11.4; PC4, 7.7)
BGM, % (PC1, 41.6; PC2, 18.7; PC3, 12.7; PC4, 12.1)

**Schist**, B

B, % (PC1, 66.2; PC2, 17.2; PC3, 11.3)
BG, % (PC1, 50.3; PC2, 19.6; PC3, 16.4)
BGM, % (PC1, 56.1; PC2, 17; PC3, 9.6)

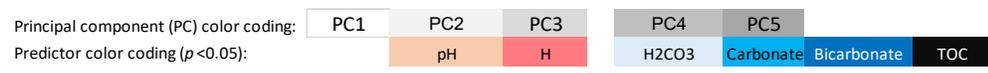

Principal component (PC) color coding: PC1 | PC2 | PC3 | PC4 | PC5
Predictor color coding ($p < 0.05$): pH | H | H2CO3 | Carbonate | Bicarbonate | TOC

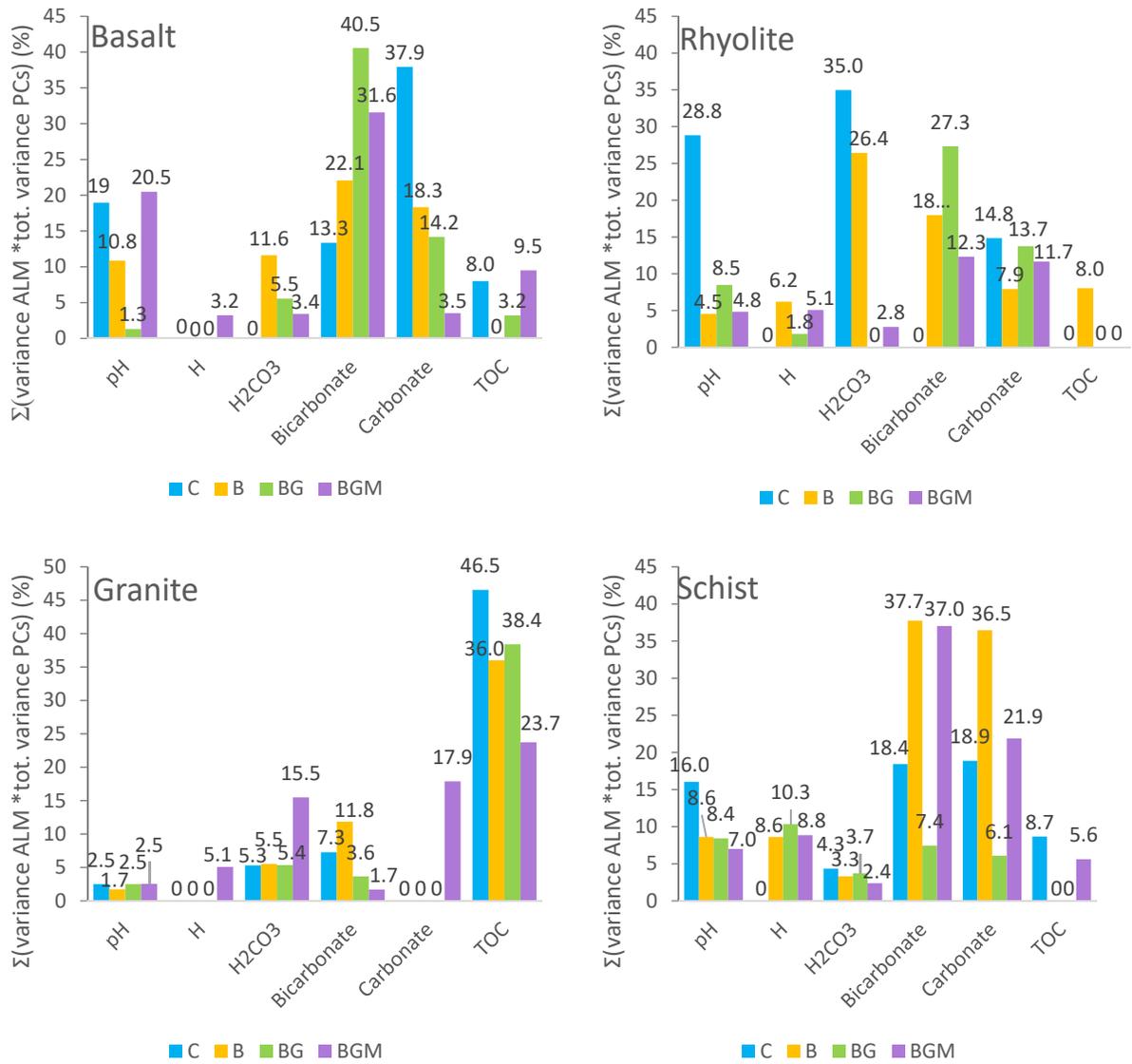

*Figure s7.* **Drivers of biotic dissolution (b).** Relative contribution of each driver to the dissolution of elements (% total variance explained by driver out of total explained by PCA; Table s3), derived from Principal Component analysis (PCA) and Automatic Linear Model (ALM). Values are significant at *p* > 95 %. Treatments: B, microbial; BG, microbial–grass; and BGM, microbial–grass–arbuscular mycorrhiza.

**SI 2.3 Vascular plant**

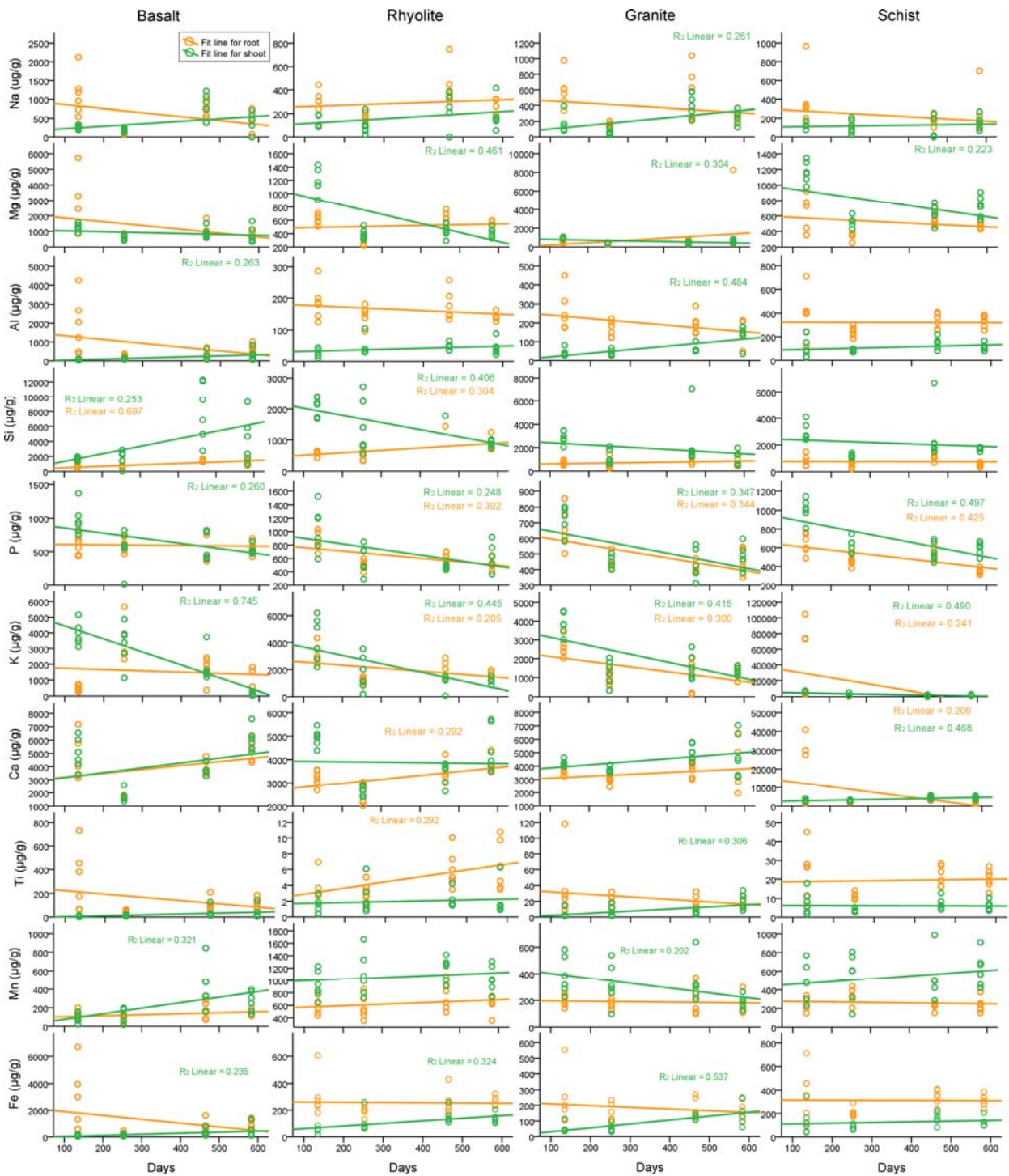

*Figure s8* **Major element content during plant development.** Time changes in major element concentrations (ug/dry biomass) in plants grown in four silicate rocks, together with their fit lines. Only $r^2 > 0.2$ is displayed.

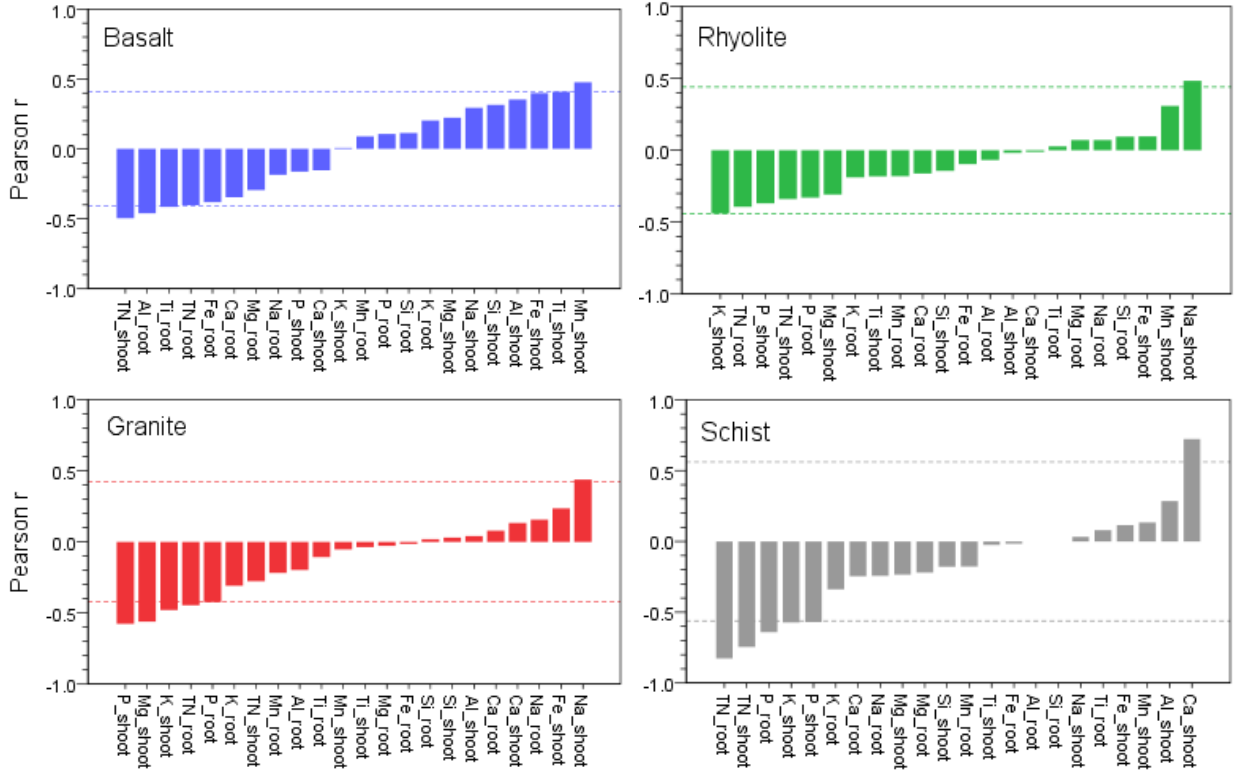

*Figure s9* **Factors limiting plant development.** Relationship between plant biomass and its element concentrations, with values above and below dash lines significant at 95% level.

## SI 2.4 Biological Signature index

*Table s4.* **Abiotic component (Z) of Biological Signature Index.** Rescaled (0-100) fractions moles element to moles total cations (summed across all sampling events) in extract : parent rock, averaged across column triplicates.

| Element | Water | | | | Exchangeable | | | | Poorly crystalline | | | |
|---|---|---|---|---|---|---|---|---|---|---|---|---|
| | Basalt | Rhyolite | Granite | Schist | Basalt | Rhyolite | Granite | Schist | Basalt | Rhyolite | Granite | Schist |
| Na | 85.14 | 9.84 | 54.20 | 42.86 | 56.30 | 2.27 | 14.75 | 69.78 | 15.15 | 51.45 | 31.16 | 38.53 |
| Mg | 5.88 | 70.95 | 67.21 | 79.22 | 21.84 | 12.84 | 37.10 | 43.26 | 18.89 | 73.83 | 76.31 | 1.13 |
| Al | 87.79 | 24.69 | 35.85 | 75.76 | 43.56 | 17.00 | 8.13 | 34.42 | 62.99 | 20.05 | 49.45 | 6.47 |
| Si | 88.90 | 66.17 | 57.12 | 9.86 | 45.58 | 63.37 | 34.07 | 83.39 | 79.04 | 42.28 | 27.20 | 19.88 |
| P | 66.82 | 19.40 | 28.85 | 53.06 | 53.12 | 86.11 | 52.30 | 24.24 | 9.08 | 16.73 | 69.23 | 46.64 |
| K | 66.88 | 29.04 | 39.90 | 20.94 | 82.47 | 15.35 | 21.66 | 8.04 | 32.63 | 58.40 | 35.64 | 89.34 |
| Ca | 11.28 | 79.92 | 45.37 | 93.25 | 47.42 | 89.50 | 75.56 | 40.05 | 33.14 | 39.97 | 51.14 | 63.81 |
| Ti | 37.59 | 29.43 | 28.99 | 21.63 | 23.01 | 19.70 | 41.51 | 26.33 | 21.50 | 12.17 | 48.96 | 16.68 |
| Mn | 9.79 | 38.99 | 63.02 | 74.92 | 54.75 | 50.78 | 26.84 | 59.35 | 12.79 | 85.00 | 0.00 | 33.33 |
| Fe | 22.79 | 31.93 | 42.04 | 60.95 | 24.16 | 13.42 | 7.30 | 18.54 | 75.06 | 54.90 | 64.02 | 10.98 |

**SI 2.5 Global projections**

*Table s5.* **Global denudation estimates.** Water denudation of major rock cations, estimated in $10^{10}$ moles/year, from four igneous lithologies under different ecosystem scenarios: C, abiotic; B, microbial; BG, microbial-vascular plant; and BGM, microbial- plant- arbuscular mycorrhiza.

| Element | Biota | Basalt | Rhyolite | Granite | Schist |
|---------|-------|--------|----------|---------|--------|
| Na | C | 73.9 | 8.64 | 27.9 | 75.5 |
|  | B | 45.2 | 18.2 | 33.2 | 116 |
|  | BG | 39.9 | 17.2 | 21.3 | 91.5 |
|  | BGM | 36.3 | 8.51 | 21.9 | 90.3 |
| Mg | C | 4.69 | 2.82 | 7.47 | 19.5 |
|  | B | 7.38 | 3.00 | 8.65 | 17.6 |
|  | BG | 7.19 | 2.65 | 6.19 | 21.3 |
|  | BGM | 6.75 | 2.97 | 6.86 | 18.5 |
| Al | C | 0.150 | 0.00868 | 0.0403 | 0.863 |
|  | B | 0.0843 | 0.0125 | 0.0569 | 0.775 |
|  | BG | 0.0694 | 0.0230 | 0.0283 | 0.467 |
|  | BGM | 0.0460 | 0.00405 | 0.0283 | 0.757 |
| Si | C | 24.1 | 13.4 | 19.2 | 65.3 |
|  | B | 17.0 | 20.6 | 23.9 | 96.8 |
|  | BG | 13.5 | 20.8 | 13.5 | 124 |
|  | BGM | 12.4 | 10.8 | 14.4 | 109 |
| P | C | 0.413 | 0.118 | 0.147 | 1.68 |
|  | B | 0.320 | 0.341 | 0.217 | 3.04 |
|  | BG | 0.239 | 0.290 | 0.139 | 0.893 |
|  | BGM | 0.178 | 0.0916 | 0.184 | 1.95 |
| K | C | 3.52 | 1.25 | 7.52 | 31.2 |
|  | B | 2.60 | 2.02 | 8.54 | 37.7 |
|  | BG | 2.31 | 2.12 | 7.20 | 54.7 |
|  | BGM | 2.10 | 1.37 | 7.17 | 51.3 |
| Ca | C | 33.6 | 12.53 | 90.7 | 78.7 |
|  | B | 30.9 | 12.3 | 89.5 | 80.6 |
|  | BG | 31.1 | 12.7 | 92.0 | 76.8 |
|  | BGM | 31.5 | 12.4 | 91.3 | 79.6 |
| Ti | C | 0.00102 | 0.000234 | 0.000412 | 0.0113 |
|  | B | 0.000630 | 0.000518 | 0.000510 | 0.0172 |
|  | BG | 0.000905 | 0.000580 | 0.000456 | 0.0130 |
|  | BGM | 0.000535 | 0.000129 | 0.000374 | 0.0219 |
| Mn | C | 0.00745 | 0.139 | 0.0577 | 0.743 |
|  | B | 0.0120 | 0.0586 | 0.0383 | 0.175 |
|  | BG | 0.00708 | 0.0891 | 0.0437 | 0.0410 |
|  | BGM | 0.0234 | 0.0619 | 0.0450 | 0.0880 |
| Fe | C | 0.00632 | 0.00328 | 0.00722 | 0.204 |
|  | B | 0.00395 | 0.00388 | 0.00649 | 0.250 |
|  | BG | 0.00483 | 0.00692 | 0.00764 | 0.136 |
|  | BGM | 0.00654 | 0.00145 | 0.00687 | 0.256 |
| **Total cations** | **C** | **140.4** | **38.9** | **153** | **273.7** |
|  | **B** | **103.5** | **56.5** | **164.1** | **353** |
|  | **BG** | **94.3** | **55.9** | **140.4** | **369.9** |
|  | **BGM** | **89.3** | **36.2** | **141.9** | **351.8** |

The contribution of abiotic and biotic processes to global denudation of terrestrial surface was calculated by normalizing total moles of element extracted during the whole experiment to global Ca + Mg estimated from large river data[2], using the equation *E. s1*. The term $i_G$ represents the global (*G*) value of element (*i*) of interest (moles * year$^{-1}$), and $i_E$ is total moles element in water from experiment *E*. Values were adjusted (multiplied) to the relative contribution (%) of the four rocks to the global lithology[3].

$$(E. s1)$$

$$i_G = \frac{i_E}{(Ca+Mg)_E} (Ca+Mg)_G$$

The contribution of different ecosystem components to global denudation was estimated from their ratio in the experiment in increasing order of complexity: GBM : GB : B : C. Results are in Table s5.

### SI References

1.  Rudnick, R. & Gao, S. Composition of the continental crust. *Treatise on Geochemistry* **3,** 1–64 (2003).

2.  Gaillardet, J., Duprre, B., Louvat, P. & Allegre, C. J. Global silicate weathering and CO2 consumption rates deduced from the chemistry of large rivers. *Chem. Geol.* **159,** 3–30 (1999).

3.  Suchet, P. A. Worldwide distribution of continental rock lithology: Implications for the atmospheric/soil CO 2 uptake by continental weathering and alkalinity river transport to the oceans. *Global Biogeochem. Cycles* **17,** (2003).

# SUPPLEMENTARY DATA

*Dataset 1* **A mass balance of abiotic and biological weathering.** Compartmentalization of major elements into shoot, root, pore water, exchangeable (ammonium acetate extract; aa), poorly crystalline (ammonium oxalate extract; ao), and unreacted rock, following a two−year weathering experiment of basalt, rhyolite, granite and schist. T0, unreacted rock; C, abiotic control; B, microbes; BG, microbes-grass; BGM, microbes-grass-arbuscular mycorrhiza. Reacted treatments are highlighted in color.

| Index | Elem Code | Rock Code | Fraction Code | Element | ROCK | Fraction | T0 N | T0 Mean %±SE | C N | C Mean %±Std. Error | B N | B Mean %±Std. Error | BG N | BG Mean %±Std. Error | BGM N | BGM Mean %±Std. Error |
|---|---|---|---|---|---|---|---|---|---|---|---|---|---|---|---|---|
| 1 | 1 | 1 | 1 | Na | Basalt | shoot | | | | | | | 3 | 0.00018±0.00018 | 3 | 0.00027±0.00007 |
| 2 | 1 | 1 | 2 | Na | Basalt | root | | | | | | | 3 | 0.00026±0.00026 | 3 | 0.00034±0.00014 |
| 3 | 1 | 1 | 3 | Na | Basalt | water | | | 3 | 0.11325±0.00235 | 3 | 0.08752±0.00046 | 3 | 0.09211±0.00019 | 3 | 0.09778±0.00683 |
| 4 | 1 | 1 | 4 | Na | Basalt | aa | 3 | 0.12844±0.001643 | 3 | 0.08226±0.00542 | 3 | 0.07701±0.00192 | 3 | 0.08256±0.00536 | 3 | 0.07346±0.00179 |
| 5 | 1 | 1 | 5 | Na | Basalt | ao | 3 | 4.86919±0.065733 | 3 | 2.16176±0.10964 | 3 | 2.41865±0.09938 | 3 | 2.30252±0.01540 | 3 | 2.57018±0.17745 |
| 6 | 1 | 1 | 6 | Na | Basalt | unreacted | 3 | 95.00238±0.06554 | 3 | 97.64273±0.11348 | 3 | 97.41683±0.09769 | 3 | 97.52237±0.01025 | 3 | 97.25796±0.18218 |
| 7 | 1 | 2 | 1 | Na | Rhyolite | shoot | | | | | | | 3 | 0.00006±0.00003 | 3 | 0.00014±0.00004 |
| 8 | 1 | 2 | 2 | Na | Rhyolite | root | | | | | | | 3 | 0.00013±0.00005 | 3 | 0.00020±0.00005 |
| 9 | 1 | 2 | 3 | Na | Rhyolite | water | | | 3 | 0.02099±0.00166 | 3 | 0.03717±0.00062 | 3 | 0.03738±0.00266 | 3 | 0.02384±0.00114 |
| 10 | 1 | 2 | 4 | Na | Rhyolite | aa | 3 | 0.04535±0.00084 | 3 | 0.02500±0.00048 | 3 | 0.04175±0.00384 | 3 | 0.05282±0.01038 | 3 | 0.02419±0.00103 |
| 11 | 1 | 2 | 5 | Na | Rhyolite | ao | 3 | 0.00884±0.000303 | 3 | 0.03410±0.00650 | 3 | 0.02470±0.00381 | 3 | 0.02425±0.00039 | 3 | 0.01755±0.00076 |
| 12 | 1 | 2 | 6 | Na | Rhyolite | unreacted | 3 | 99.94581±0.00111 | 3 | 99.91991±0.00532 | 3 | 99.89637±0.00209 | 3 | 99.88536±0.01254 | 3 | 99.93407±0.00219 |
| 13 | 1 | 3 | 1 | Na | Granite | shoot | | | | | | | 3 | 0.00012±0.00001 | 3 | 0.00012±0.00001 |
| 14 | 1 | 3 | 2 | Na | Granite | root | | | | | | | 3 | 0.00018±0.00005 | 3 | 0.00019±0.00001 |
| 15 | 1 | 3 | 3 | Na | Granite | water | | | 3 | 0.03338±0.00085 | 3 | 0.03079±0.00122 | 3 | 0.03070±0.00112 | 3 | 0.03418±0.00156 |
| 16 | 1 | 3 | 4 | Na | Granite | aa | 3 | 0.05560±0.000693 | 3 | 0.01184±0.00127 | 3 | 0.01183±0.00077 | 3 | 0.01258±0.00252 | 3 | 0.01734±0.00243 |
| 17 | 1 | 3 | 5 | Na | Granite | ao | 3 | 0.02427±0.000773 | 3 | 0.02325±0.00078 | 3 | 0.02087±0.00116 | 3 | 0.02309±0.00134 | 3 | 0.03723±0.00248 |
| 18 | 1 | 3 | 6 | Na | Granite | unreacted | 3 | 99.92012±0.00038 | 3 | 99.93152±0.00050 | 3 | 99.93651±0.00153 | 3 | 99.93334±0.00164 | 3 | 99.91093±0.00487 |
| 19 | 1 | 4 | 1 | Na | Schist | shoot | | | | | | | 3 | 0.00004±0.00003 | 3 | 0.00134±0.00024 |
| 20 | 1 | 4 | 2 | Na | Schist | root | | | | | | | 3 | 0.00210±0.00089 | 3 | 0.00456±0.00055 |
| 21 | 1 | 4 | 3 | Na | Schist | water | | | 3 | 0.24088±0.01752 | 3 | 0.21728±0.00376 | 3 | 0.31277±0.01020 | 3 | 0.22695±0.01283 |
| 22 | 1 | 4 | 4 | Na | Schist | aa | 3 | 0.38854±0.01084 | 3 | 0.06822±0.00955 | 3 | 0.06038±0.01441 | 3 | 0.07227±0.00416 | 3 | 0.03840±0.00149 |
| 23 | 1 | 4 | 5 | Na | Schist | ao | 3 | 0.02927±0.00953 | 3 | 0.16037±0.02824 | 3 | 0.19489±0.01331 | 3 | 0.19972±0.01410 | 3 | 0.16741±0.00860 |
| 24 | 1 | 4 | 6 | Na | Schist | unreacted | 3 | 99.58219±0.01193 | 3 | 99.53053±0.02892 | 3 | 99.52745±0.00490 | 3 | 99.41222±0.00890 | 3 | 99.56150±0.01628 |
| 25 | 2 | 1 | 1 | Mg | Basalt | shoot | | | | | | | 3 | 0.00010±0.00004 | 3 | 0.00022±0.00009 |
| 26 | 2 | 1 | 2 | Mg | Basalt | root | | | | | | | 3 | 0.00020±0.00006 | 3 | 0.00032±0.00001 |
| 27 | 2 | 1 | 3 | Mg | Basalt | water | | | 3 | 0.00252±0.00012 | 3 | 0.00501±0.00020 | 3 | 0.00581±0.00038 | 3 | 0.00637±0.00069 |
| 28 | 2 | 1 | 4 | Mg | Basalt | aa | 3 | 0.05596±0.00084 | 3 | 0.04114±0.00069 | 3 | 0.04658±0.00163 | 3 | 0.04610±0.00262 | 3 | 0.04543±0.00240 |
| 29 | 2 | 1 | 5 | Mg | Basalt | ao | 3 | 2.45053±0.014123 | 3 | 1.34080±0.06703 | 3 | 1.47781±0.05354 | 3 | 1.40765±0.00800 | 3 | 1.56282±0.03880 |
| 30 | 2 | 1 | 6 | Mg | Basalt | unreacted | 3 | 98.61554±0.06756 | 3 | 98.47059±0.05339 | 3 | 98.54014±0.06690 | 3 | 98.38483±0.09051 | | |
| 31 | 2 | 2 | 1 | Mg | Rhyolite | shoot | | | | | | | 3 | 0.00460±0.00119 | 3 | 0.00727±0.00137 |
| 32 | 2 | 2 | 2 | Mg | Rhyolite | root | | | | | | | 3 | 0.00663±0.00138 | 3 | 0.01173±0.00229 |
| 33 | 2 | 2 | 3 | Mg | Rhyolite | water | | | 3 | 0.19012±0.02575 | 3 | 0.17099±0.02302 | 3 | 0.15907±0.02735 | 3 | 0.23078±0.00965 |
| 34 | 2 | 2 | 4 | Mg | Rhyolite | aa | 3 | 6.15617±0.194023 | 3 | 5.62224±0.11237 | 3 | 6.63579±0.07866 | 3 | 6.29286±0.16976 | 3 | 5.54934±0.07184 |
| 35 | 2 | 2 | 5 | Mg | Rhyolite | ao | 3 | 1.34817±0.041633 | 3 | 1.41582±0.06029 | 3 | 1.20626±0.02766 | 3 | 1.21219±0.01538 | 3 | 1.24135±0.03234 |
| 36 | 2 | 2 | 6 | Mg | Rhyolite | unreacted | 3 | 92.49566±0.22592 | 3 | 92.77183±0.13302 | 3 | 91.98697±0.10224 | 3 | 92.32465±0.16132 | 3 | 92.95953±0.10267 |
| 37 | 2 | 3 | 1 | Mg | Granite | shoot | | | | | | | 3 | 0.00130±0.00012 | 3 | 0.00171±0.00021 |
| 38 | 2 | 3 | 2 | Mg | Granite | root | | | | | | | 3 | 0.01225±0.01019 | 3 | 0.00226±0.00014 |
| 39 | 2 | 3 | 3 | Mg | Granite | water | | | 3 | 0.05373±0.00098 | 3 | 0.04818±0.00231 | 3 | 0.05354±0.00423 | 3 | 0.06439±0.00622 |
| 40 | 2 | 3 | 4 | Mg | Granite | aa | 3 | 0.38638±0.00447 | 3 | 0.15777±0.00388 | 3 | 0.17141±0.00273 | 3 | 0.12463±0.01057 | 3 | 0.13064±0.00476 |
| 41 | 2 | 3 | 5 | Mg | Granite | ao | 3 | 0.73396±0.04131 | 3 | 0.13401±0.00970 | 3 | 0.12042±0.00824 | 3 | 0.12984±0.00457 | 3 | 0.14108±0.01102 |
| 42 | 2 | 3 | 6 | Mg | Granite | unreacted | 3 | 98.87965±0.04546 | 3 | 99.65449±0.01445 | 3 | 99.65999±0.02852 | 3 | 99.67844±0.00599 | 3 | 99.65994±0.01118 |
| 43 | 2 | 4 | 1 | Mg | Schist | shoot | | | | | | | 3 | 0.00073±0.00013 | 3 | 0.00064±0.00003 |
| 44 | 2 | 4 | 2 | Mg | Schist | root | | | | | | | 3 | 0.00080±0.00007 | 3 | 0.00078±0.00005 |
| 45 | 2 | 4 | 3 | Mg | Schist | water | | | 3 | 0.00823±0.00063 | 3 | 0.00433±0.00030 | 3 | 0.00964±0.00113 | 3 | 0.00619±0.00072 |
| 46 | 2 | 4 | 4 | Mg | Schist | aa | 3 | 0.08788±0.00480 | 3 | 0.03608±0.00096 | 3 | 0.03607±0.00215 | 3 | 0.05475±0.00262 | 3 | 0.04483±0.00375 |
| 47 | 2 | 4 | 5 | Mg | Schist | ao | 3 | 0.08629±0.00363 | 3 | 0.01060±0.00049 | 3 | 0.06608±0.05469 | 3 | 0.03597±0.00166 | 3 | 0.02552±0.00124 |
| 48 | 2 | 4 | 6 | Mg | Schist | unreacted | 3 | 99.94508±0.00134 | 3 | 99.89352±0.05687 | 3 | 99.89812±0.00138 | 3 | 99.92240±0.01161 | | |
| 49 | 3 | 1 | 1 | Al | Basalt | shoot | | | | | | | 3 | 0.00002±0.00000 | 3 | 0.00004±0.00004 |
| 50 | 3 | 1 | 2 | Al | Basalt | root | | | | | | | 3 | 0.00012±0.00003 | 3 | 0.00018±0.00000 |
| 51 | 3 | 1 | 3 | Al | Basalt | water | | | 3 | 0.00007±0.00000 | 3 | 0.00005±0.00001 | 3 | 0.00005±0.00000 | 3 | 0.00005±0.00000 |
| 52 | 3 | 1 | 4 | Al | Basalt | aa | 3 | 0.04061±0.00048 | 3 | 0.04537±0.00067 | 3 | 0.04631±0.00070 | 3 | 0.04746±0.00086 | 3 | 0.04794±0.00061 |
| 53 | 3 | 1 | 5 | Al | Basalt | ao | 3 | 2.67454±0.017293 | 3 | 1.56041±0.00687 | 3 | 1.58567±0.19502 | 3 | 1.54218±0.02793 | 3 | 1.74974±0.15727 |
| 54 | 3 | 1 | 6 | Al | Basalt | unreacted | 3 | 97.28484±0.01730 | 3 | 98.39415±0.06119 | 3 | 98.36796±0.19527 | 3 | 98.41016±0.02871 | 3 | 98.20203±0.19196 |
| 55 | 3 | 2 | 1 | Al | Rhyolite | shoot | | | | | | | 3 | 0.00001±0.00001 | 3 | 0.00001±0.00000 |
| 56 | 3 | 2 | 2 | Al | Rhyolite | root | | | | | | | 3 | 0.00003±0.00001 | 3 | 0.00005±0.00001 |
| 57 | 3 | 2 | 3 | Al | Rhyolite | water | | | 3 | 0.00001±0.00000 | 3 | 0.00001±0.00000 | 3 | 0.00003±0.00001 | 3 | 0.00001±0.00000 |
| 58 | 3 | 2 | 4 | Al | Rhyolite | aa | 3 | 0.01926±0.00038 | 3 | 0.02264±0.00093 | 3 | 0.03327±0.00136 | 3 | 0.03030±0.00204 | 3 | 0.02320±0.00036 |
| 59 | 3 | 2 | 5 | Al | Rhyolite | ao | 3 | 0.16632±0.00040 | 3 | 0.16221±0.00183 | 3 | 0.22756±0.00378 | 3 | 0.22135±0.01934 | 3 | 0.16690±0.00148 |
| 60 | 3 | 2 | 6 | Al | Rhyolite | unreacted | 3 | 99.81442±0.00476 | 3 | 99.81514±0.00267 | 3 | 99.73916±0.00502 | 3 | 99.74682±0.02093 | 3 | 99.80983±0.00183 |
| 61 | 3 | 3 | 1 | Al | Granite | shoot | | | | | | | 3 | 0.00003±0.00000 | 3 | 0.00001±0.00000 |
| 62 | 3 | 3 | 2 | Al | Granite | root | | | | | | | 3 | 0.00003±0.00001 | 3 | 0.00005±0.00002 |
| 63 | 3 | 3 | 3 | Al | Granite | water | | | 3 | 0.00002±0.00000 | 3 | 0.00002±0.00000 | 3 | 0.00002±0.00000 | 3 | 0.00002±0.00000 |

| # | | | | El | Rock | Sample | M1 | M2 | M3 | M4 | M5 |
|---|---|---|---|---|---|---|---|---|---|---|---|
| 64 | 3 | 3 | 4 | Al | Granite | aa | 3 0.01086±0.00028 | 3 0.00506±0.00008 | 3 0.00762±0.00275 | 3 0.00531±0.00007 | 3 0.00640±0.00016 |
| 65 | 3 | 3 | 5 | Al | Granite | ao | 3 0.10134±0.00220 | 3 0.04227±0.00169 | 3 0.03872±0.00188 | 3 0.04468±0.00130 | 3 0.04607±0.00242 |
| 66 | 3 | 3 | 6 | Al | Granite | unextracted | 3 99.88780±0.00242 | 3 99.95265±0.00172 | 3 99.95364±0.00105 | 3 99.94994±0.00123 | 3 99.94745±0.00257 |
| 67 | 3 | 4 | 1 | Al | Schist | shoot | | | | 3 0.00002±0.00000 | 3 0.00002±0.00001 |
| 68 | 3 | 4 | 2 | Al | Schist | root | | | | 3 0.00011±0.00001 | 3 0.00005±0.00000 |
| 69 | 3 | 4 | 3 | Al | Schist | water | | 3 0.00007±0.00001 | 3 0.00004±0.00000 | 3 0.00004±0.00000 | 3 0.00005±0.00000 |
| 70 | 3 | 4 | 4 | Al | Schist | aa | 3 0.01464±0.00027 | 3 0.00369±0.00048 | 3 0.00366±0.00057 | 3 0.00691±0.00022 | 3 0.00487±0.00102 |
| 71 | 3 | 4 | 5 | Al | Schist | ao | 3 0.05839±0.00154 | 3 0.01129±0.00014 | 3 0.01262±0.00065 | 3 0.04006±0.00128 | 3 0.02597±0.00763 |
| 72 | 3 | 4 | 6 | Al | Schist | unextracted | 3 99.92697±0.00166 | 3 99.98494±0.00036 | 3 99.98368±0.00118 | 3 99.95286±0.00145 | 3 99.96899±0.00866 |
| 73 | 4 | 1 | 1 | Si | Basalt | shoot | | | | 3 0.00008±0.00002 | 3 0.00013±0.00000 |
| 74 | 4 | 1 | 2 | Si | Basalt | root | | | | 3 0.00010±0.00002 | 3 0.00013±0.00001 |
| 75 | 4 | 1 | 3 | Si | Basalt | water | | 3 0.00461±0.00009 | 3 0.00411±0.00018 | 3 0.00387±0.00030 | 3 0.00415±0.00004 |
| 76 | 4 | 1 | 4 | Si | Basalt | aa | 3 0.03015±0.00203 | 3 0.01092±0.00048 | 3 0.01059±0.00055 | 3 0.01180±0.00062 | 3 0.01176±0.00065 |
| 77 | 4 | 1 | 5 | Si | Basalt | ao | 3 2.32300±0.04205 | 3 1.10731±0.04107 | 3 1.10846±0.15526 | 3 1.09096±0.02890 | 3 1.24052±0.13108 |
| 78 | 4 | 1 | 6 | Si | Basalt | unextracted | 3 97.64685±0.04225 | 3 98.87716±0.04141 | 3 98.87684±0.15565 | 3 98.89318±0.02905 | 3 98.74302±0.13114 |
| 79 | 4 | 2 | 1 | Si | Rhyolite | shoot | | | | 3 0.00004±0.00001 | 3 0.00005±0.00001 |
| 80 | 4 | 2 | 2 | Si | Rhyolite | root | | | | 3 0.00005±0.00002 | 3 0.00007±0.00001 |
| 81 | 4 | 2 | 3 | Si | Rhyolite | water | | 3 0.00410±0.00019 | 3 0.00528±0.00016 | 3 0.00567±0.00020 | 3 0.00379±0.00010 |
| 82 | 4 | 2 | 4 | Si | Rhyolite | aa | 3 0.01463±0.00026 | 3 0.00412±0.00033 | 3 0.00515±0.00021 | 3 0.00441±0.00022 | 3 0.00292±0.00035 |
| 83 | 4 | 2 | 5 | Si | Rhyolite | ao | 3 0.02190±0.00042 | 3 0.02663±0.00053 | 3 0.02792±0.00022 | 3 0.02842±0.00072 | 3 0.02625±0.00078 |
| 84 | 4 | 2 | 6 | Si | Rhyolite | unextracted | 3 99.96347±0.00068 | 3 99.96515±0.00046 | 3 99.96165±0.00029 | 3 99.96141±0.00070 | 3 99.96692±0.00096 |
| 85 | 4 | 3 | 1 | Si | Granite | shoot | | | | 3 0.00003±0.00001 | 3 0.00007±0.00000 |
| 86 | 4 | 3 | 2 | Si | Granite | root | | | | 3 0.00007±0.00003 | 3 0.00002±0.00000 |
| 87 | 4 | 3 | 3 | Si | Granite | water | | 3 0.00237±0.00005 | 3 0.00229±0.00006 | 3 0.00201±0.00008 | 3 0.00232±0.00005 |
| 88 | 4 | 3 | 4 | Si | Granite | aa | 3 0.01210±0.00193 | 3 0.00034±0.00005 | 3 0.00041±0.00018 | 1 0.0054 | 3 0.00087±0.00005 |
| 89 | 4 | 3 | 5 | Si | Granite | ao | 3 0.02350±0.00037 | 3 0.01479±0.00012 | 3 0.01474±0.00111 | 3 0.01574±0.00006 | 3 0.01700±0.00075 |
| 90 | 4 | 3 | 6 | Si | Granite | unextracted | 3 99.96440±0.00041 | 3 99.98250±0.00112 | 3 99.98256±0.00097 | 3 99.98197±0.00010 | 3 99.97970±0.00081 |
| 91 | 4 | 4 | 1 | Si | Schist | shoot | | | | 3 0.00009±0.00001 | 3 0.00009±0.00001 |
| 92 | 4 | 4 | 2 | Si | Schist | root | | | | 3 0.00005±0.00001 | 3 0.00006±0.00001 |
| 93 | 4 | 4 | 3 | Si | Schist | water | | 3 0.00182±0.00011 | 3 0.00158±0.00002 | 3 0.00240±0.00004 | 3 0.00042±0.00042 |
| 94 | 4 | 4 | 4 | Si | Schist | aa | 3 0.01447±0.00012 | 3 0.00183±0.00028 | 3 0.00129±0.00017 | 3 0.00116±0.00011 | 3 0.00036±0.00018 |
| 95 | 4 | 4 | 5 | Si | Schist | ao | 3 0.01495±0.00035 | 3 0.00853±0.00051 | 3 0.01189±0.00289 | 3 0.01231±0.00028 | 3 0.01042±0.00106 |
| 96 | 4 | 4 | 6 | Si | Schist | unextracted | 3 99.97058±0.00066 | 3 99.98782±0.00036 | 3 99.98524±0.00275 | 3 99.98720±0.00040 | 3 99.98667±0.00136 |
| 97 | 5 | 1 | 1 | P | Basalt | shoot | | | | 3 0.00213±0.00054 | 3 0.00273±0.00031 |
| 98 | 5 | 1 | 2 | P | Basalt | root | | | | 3 0.00317±0.00062 | 3 0.00456±0.00073 |
| 99 | 5 | 1 | 3 | P | Basalt | water | | 3 0.00664±0.00058 | 3 0.00651±0.00065 | 3 0.00579±0.00016 | 3 0.00503±0.00020 |
| 100 | 5 | 1 | 4 | P | Basalt | aa | 3 0.72124±0.00700 | 3 0.47358±0.00841 | 3 0.45043±0.00535 | 3 0.46298±0.01700 | 3 0.46713±0.01753 |
| 101 | 5 | 1 | 5 | P | Basalt | ao | 3 9.55321±0.19719 | 3 5.33289±0.25761 | 3 6.08242±0.13371 | 3 5.80246±0.07339 | 3 6.39989±0.38527 |
| 102 | 5 | 1 | 6 | P | Basalt | unextracted | 3 89.72555±0.20249 | 3 94.18690±0.26379 | 3 93.46064±0.12902 | 3 93.72349±0.08904 | 3 93.12067±0.40114 |
| 103 | 5 | 2 | 1 | P | Rhyolite | shoot | | | | 3 0.02335±0.00667 | 3 0.03896±0.01457 |
| 104 | 5 | 2 | 2 | P | Rhyolite | root | | | | 3 0.02664±0.00645 | 3 0.03355±0.00697 |
| 105 | 5 | 2 | 3 | P | Rhyolite | water | | 3 0.03347±0.00368 | 3 0.08136±0.00612 | 3 0.07399±0.01308 | 3 0.02978±0.00340 |
| 106 | 5 | 2 | 4 | P | Rhyolite | aa | 3 1.58164±0.16217 | 3 1.18135±0.08198 | 3 0.42494±0.03161 | 3 0.71397±0.05275 | 3 0.72043±0.16297 |
| 107 | 5 | 2 | 5 | P | Rhyolite | ao | 3 17.48938±0.53915 | 3 13.73536±0.58663 | 3 15.91596±0.52300 | 3 15.19889±0.12052 | 3 14.26216±0.89916 |
| 108 | 5 | 2 | 6 | P | Rhyolite | unextracted | 3 80.92898±0.38550 | 3 85.04981±0.66301 | 3 83.57775±0.50358 | 3 83.96317±0.12736 | 3 84.91512±1.00831 |
| 109 | 5 | 3 | 1 | P | Granite | shoot | | | | 3 0.01471±0.00294 | 3 0.01471±0.00086 |
| 110 | 5 | 3 | 2 | P | Granite | root | | | | 3 0.02066±0.00591 | 3 0.02273±0.00266 |
| 111 | 5 | 3 | 3 | P | Granite | water | | 3 0.01523±0.00307 | 3 0.01733±0.00298 | 3 0.01725±0.00194 | 3 0.02485±0.00383 |
| 112 | 5 | 3 | 4 | P | Granite | aa | 3 1.49371±0.09345 | 3 0.39758±0.07281 | 3 0.31223±0.07732 | 3 0.27763±0.01170 | 3 0.41283±0.02592 |
| 113 | 5 | 3 | 5 | P | Granite | ao | 3 16.94637±0.98793 | 3 11.98125±0.19107 | 3 11.32337±0.67795 | 3 11.84283±0.37940 | 3 10.26303±0.53384 |
| 114 | 5 | 3 | 6 | P | Granite | unextracted | 3 81.55991±0.96629 | 3 87.60595±0.13918 | 3 88.34707±0.62986 | 3 87.82691±0.38083 | 3 89.26185±0.51160 |
| 115 | 5 | 4 | 1 | P | Schist | shoot | | | | 3 0.05654±0.00904 | 3 0.05293±0.00556 |
| 116 | 5 | 4 | 2 | P | Schist | root | | | | 3 0.05896±0.00933 | 3 0.05845±0.00311 |
| 117 | 5 | 4 | 3 | P | Schist | water | | 3 0.08861±0.01416 | 3 0.09396±0.00253 | 3 0.05060±0.00524 | 3 0.07977±0.01885 |
| 118 | 5 | 4 | 4 | P | Schist | aa | 3 2.10413±0.04803 | 3 0.18292±0.09990 | 3 0.34955±0.05436 | 3 0.82793±0.04172 | 3 0.26291±0.10255 |
| 119 | 5 | 4 | 5 | P | Schist | ao | 3 12.17993±0.89321 | 3 18.53481±0.65225 | 3 20.33146±0.60153 | 3 21.07635±1.52255 | 3 22.48031±0.82520 |
| 120 | 5 | 4 | 6 | P | Schist | unextracted | 3 85.71594±0.89486 | 3 81.19365±0.71867 | 3 79.22502±0.56107 | 3 77.92963±1.50107 | 3 77.06563±0.74246 |
| 121 | 6 | 1 | 1 | K | Basalt | shoot | | | | 3 0.00020±0.00013 | 3 0.00036±0.00017 |
| 122 | 6 | 1 | 2 | K | Basalt | root | | | | 3 0.00201±0.00187 | 3 0.00304±0.00205 |
| 123 | 6 | 1 | 3 | K | Basalt | water | | 3 0.03243±0.00027 | 3 0.03038±0.00019 | 3 0.03208±0.00124 | 3 0.03405±0.00282 |
| 124 | 6 | 1 | 4 | K | Basalt | aa | 3 0.32224±0.03673 | 3 0.31966±0.01202 | 3 0.29020±0.00289 | 3 0.29870±0.01177 | 3 0.28833±0.00756 |
| 125 | 6 | 1 | 5 | K | Basalt | ao | 3 9.76332±0.04623 | 3 6.09482±0.03491 | 3 6.47188±0.12948 | 3 6.43077±0.16314 | 3 6.90077±0.36759 |
| 126 | 6 | 1 | 6 | K | Basalt | unextracted | 3 89.91445±0.00095 | 3 93.55309±0.32482 | 3 93.20753±0.12788 | 3 93.23624±0.15957 | 3 92.77345±0.37526 |
| 127 | 6 | 2 | 1 | K | Rhyolite | shoot | | | | 3 0.00052±0.00010 | 3 0.00061±0.00015 |
| 128 | 6 | 2 | 2 | K | Rhyolite | root | | | | 3 0.00080±0.00023 | 3 0.00114±0.00034 |
| 129 | 6 | 2 | 3 | K | Rhyolite | water | | 3 0.00473±0.00029 | 3 0.00642±0.00028 | 3 0.00716±0.00023 | 3 0.00595±0.00024 |
| 130 | 6 | 2 | 4 | K | Rhyolite | aa | 3 0.08869±0.00149 | 3 0.08515±0.00465 | 3 0.12736±0.00456 | 3 0.12447±0.01199 | 3 0.08417±0.00075 |
| 131 | 6 | 2 | 5 | K | Rhyolite | ao | 3 0.57683±0.01108 | 3 0.54212±0.01141 | 3 0.49889±0.00516 | 3 0.51999±0.00110 | 3 0.51265±0.00207 |
| 132 | 6 | 2 | 6 | K | Rhyolite | unextracted | 3 99.33448±0.01256 | 3 99.36800±0.01441 | 3 99.36733±0.00091 | 3 99.34707±0.01181 | 3 99.39548±0.00322 |
| 133 | 6 | 3 | 1 | K | Granite | shoot | | | | 3 0.00046±0.00002 | 3 0.00059±0.00003 |
| 134 | 6 | 3 | 2 | K | Granite | root | | | | 3 0.00072±0.00026 | 3 0.00075±0.00014 |
| 135 | 6 | 3 | 3 | K | Granite | water | | 3 0.01243±0.00039 | 3 0.01092±0.00012 | 3 0.01433±0.00118 | 3 0.01545±0.00088 |
| 136 | 6 | 3 | 4 | K | Granite | aa | 3 0.04199±0.00116 | 3 0.01984±0.00093 | 3 0.01971±0.00111 | 3 0.01995±0.00078 | 3 0.02458±0.00167 |
| 137 | 6 | 3 | 5 | K | Granite | ao | 3 0.59933±0.00531 | 3 0.57797±0.00589 | 3 0.57481±0.01405 | 3 0.57082±0.00684 | 3 0.58912±0.01037 |
| 138 | 6 | 3 | 6 | K | Granite | unextracted | 3 99.35868±0.00644 | 3 99.38977±0.00660 | 3 99.39456±0.01338 | 3 99.39372±0.00702 | 3 99.36951±0.01211 |
| 139 | 6 | 4 | 1 | K | Schist | shoot | | | | 3 0.00053±0.00016 | 3 0.00056±0.00015 |
| 140 | 6 | 4 | 2 | K | Schist | root | | | | 3 0.00064±0.00017 | 3 0.00052±0.00011 |
| 141 | 6 | 4 | 3 | K | Schist | water | | 3 0.00653±0.00024 | 3 0.00462±0.00006 | 3 0.01227±0.00067 | 3 0.00852±0.00123 |
| 142 | 6 | 4 | 4 | K | Schist | aa | 3 0.03562±0.01003 | 3 0.01083±0.00082 | 3 0.01133±0.00105 | 3 0.02944±0.00053 | 3 0.01927±0.00468 |
| 143 | 6 | 4 | 5 | K | Schist | ao | 3 0.39492±0.00753 | 3 0.36982±0.00464 | 3 0.37464±0.00340 | 3 0.38004±0.00319 | 3 0.36469±0.00474 |
| 144 | 6 | 4 | 6 | K | Schist | unextracted | 3 99.56946±0.00850 | 3 99.61281±0.00530 | 3 99.60941±0.00208 | 3 99.57708±0.00360 | 3 99.60645±0.01041 |

| # | | | | El | Rock | Trt | | | | | |
|---|---|---|---|---|---|---|---|---|---|---|---|
| 145 | 7 | 1 | 1 | Ca | Basalt | shoot | | | | 3 0.00106±0.00026 | 3 0.00122±0.00025 |
| 146 | 7 | 1 | 2 | Ca | Basalt | root | | | | 3 0.00139±0.00016 | 3 0.00164±0.00014 |
| 147 | 7 | 1 | 3 | Ca | Basalt | water | | 3 0.02965±0.00118 | 3 0.03446±0.00195 | 3 0.04121±0.00271 | 3 0.04891±0.00638 |
| 148 | 7 | 1 | 4 | Ca | Basalt | aa | 3 0.90106±0.05545 | 3 0.43142±0.03720 | 3 0.44636±0.03804 | 3 0.45467±0.07379 | 3 0.42138±0.02749 |
| 149 | 7 | 1 | 5 | Ca | Basalt | ao | 3 0.06656±0.00613 | 3 0.08842±0.00654 | 3 0.09328±0.00176 | 3 0.08107±0.00867 | 3 0.09450±0.00249 |
| 150 | 7 | 1 | 6 | Ca | Basalt | unextracted | 3 99.03238±0.06080 | 3 99.45052±0.04300 | 3 99.42590±0.03678 | 3 99.40080±0.07965 | 3 99.43235±0.04453 |
| 151 | 7 | 2 | 1 | Ca | Rhyolite | shoot | | | | 3 0.00775±0.00158 | 3 0.01481±0.00358 |
| 152 | 7 | 2 | 2 | Ca | Rhyolite | root | | | | 3 0.00945±0.00139 | 3 0.01391±0.00302 |
| 153 | 7 | 2 | 3 | Ca | Rhyolite | water | | 3 0.24353±0.03068 | 3 0.20271±0.01991 | 3 0.21935±0.04183 | 3 0.27708±0.01080 |
| 154 | 7 | 2 | 4 | Ca | Rhyolite | aa | 3 11.84950±0.29141 | 7.98953±0.14047 | 3 8.23971±0.29495 | 7.52836±0.27066 | 6.66611±0.21646 |
| 155 | 7 | 2 | 5 | Ca | Rhyolite | ao | 3 0.73299±0.05768 | 1.39833±0.65383 | 3 0.61787±0.00648 | 0.69582±0.01796 | 3 0.73227±0.08620 |
| 156 | 7 | 2 | 6 | Ca | Rhyolite | unextracted | 3 87.41752±0.31016 | 3 90.36862±0.79803 | 3 90.93970±0.28078 | 91.53927±0.22747 | 3 92.29582±0.29402 |
| 157 | 7 | 3 | 1 | Ca | Granite | shoot | | | | 3 0.00318±0.00052 | 3 0.00437±0.00045 |
| 158 | 7 | 3 | 2 | Ca | Granite | root | | | | 3 0.00608±0.00219 | 3 0.00362±0.00055 |
| 159 | 7 | 3 | 3 | Ca | Granite | water | | 3 0.31430±0.00880 | 3 0.24020±0.01025 | 3 0.38381±0.03786 | 3 0.41311±0.05115 |
| 160 | 7 | 3 | 4 | Ca | Granite | aa | 3 6.14749±0.22641 | 2.70519±0.32088 | 3 2.36068±0.06770 | 2.09349±0.10424 | 3 2.42491±0.27476 |
| 161 | 7 | 3 | 5 | Ca | Granite | ao | 3 0.07811±0.01566 | 3 0.23199±0.01610 | 3 0.19313±0.00503 | 0.21078±0.00736 | 3 0.29555±0.00544 |
| 162 | 7 | 3 | 6 | Ca | Granite | unextracted | 3 93.77440±0.23861 | 3 96.74852±0.32763 | 3 97.20599±0.08044 | 97.30266±0.08601 | 3 96.85845±0.23597 |
| 163 | 7 | 4 | 1 | Ca | Schist | shoot | | | | 3 0.08203±0.01681 | 3 0.07139±0.01304 |
| 164 | 7 | 4 | 2 | Ca | Schist | root | | | | 3 0.07403±0.01543 | 3 0.06613±0.01219 |
| 165 | 7 | 4 | 3 | Ca | Schist | water | | 3 1.05398±0.05584 | 3 0.63133±0.02580 | 3 1.10283±0.09604 | 3 0.84125±0.04151 |
| 166 | 7 | 4 | 4 | Ca | Schist | aa | 3 24.98316±5.40112 | 6.68149±0.13003 | 3 6.59238±0.07817 | 8.76629±0.35816 | 9.80685±1.71631 |
| 167 | 7 | 4 | 5 | Ca | Schist | ao | 3 2.73797±1.47272 | 3 3.40444±0.27740 | 3 3.60663±0.10499 | 3.84416±0.09955 | 3 3.39673±0.02608 |
| 168 | 7 | 4 | 6 | Ca | Schist | unextracted | 3 72.27887±4.94515 | 3 88.86009±0.38736 | 3 89.16966±0.11845 | 86.13066±0.39176 | 3 85.81765±1.80142 |
| 169 | 8 | 1 | 1 | Ti | Basalt | shoot | | | | 3 0.00001±0.00000 | 3 0.00006±0.00006 |
| 170 | 8 | 1 | 2 | Ti | Basalt | root | | | | 3 0.00020±0.00003 | 3 0.00027±0.00004 |
| 171 | 8 | 1 | 3 | Ti | Basalt | water | | 3 0.00001±0.00000 | 3 0.00001±0.00000 | 3 0.00001±0.00000 | 3 0.00001±0.00000 |
| 172 | 8 | 1 | 4 | Ti | Basalt | aa | 3 0.00429±0.00025 | 3 0.00233±0.00011 | 3 0.00287±0.00011 | 3 0.00291±0.00020 | 3 0.00292±0.00025 |
| 173 | 8 | 1 | 5 | Ti | Basalt | ao | 3 6.89495±0.11438 | 4.23606±0.16703 | 3 4.70033±0.16606 | 4.40690±0.05330 | 3 4.83926±0.30263 |
| 174 | 8 | 1 | 6 | Ti | Basalt | unextracted | 3 93.10076±0.11437 | 3 95.76160±0.16698 | 3 95.29680±0.16600 | 95.58996±0.05336 | 3 95.15747±0.30324 |
| 175 | 8 | 2 | 1 | Ti | Rhyolite | shoot | | | | 3 0.00003±0.00002 | 3 0.00001±0.00000 |
| 176 | 8 | 2 | 2 | Ti | Rhyolite | root | | | | 3 0.00008±0.00002 | 3 0.00001±0.00000 |
| 177 | 8 | 2 | 3 | Ti | Rhyolite | water | | 3 0.00002±0.00000 | 3 0.00003±0.00000 | 3 0.00004±0.00001 | 3 0.00001±0.00000 |
| 178 | 8 | 2 | 4 | Ti | Rhyolite | aa | 3 0.00083±0.00011 | 3 0.00107±0.00065 | 3 0.00235±0.00178 | 3 0.00024±0.00014 | 3 0.00026±0.00015 |
| 179 | 8 | 2 | 5 | Ti | Rhyolite | ao | 3 0.04274±0.01005 | 3 0.05026±0.04366 | 3 0.11457±0.01319 | 0.09692±0.01639 | 3 0.05434±0.00035 |
| 180 | 8 | 2 | 6 | Ti | Rhyolite | unextracted | 3 99.95643±0.00099 | 3 99.94866±0.00311 | 3 99.88305±0.01257 | 99.90269±0.01655 | 3 99.94532±0.00048 |
| 181 | 8 | 3 | 1 | Ti | Granite | shoot | | | | 3 0.00012±0.00002 | 3 0.00008±0.00000 |
| 182 | 8 | 3 | 2 | Ti | Granite | root | | | | 3 0.00016±0.00005 | 3 0.00017±0.00003 |
| 183 | 8 | 3 | 3 | Ti | Granite | water | | 3 0.00001±0.00000 | 3 0.00001±0.00000 | 3 0.00003±0.00001 | 3 0.00003±0.00000 |
| 184 | 8 | 3 | 4 | Ti | Granite | aa | 3 0.00446±0.00025 | 3 0.00150±0.00013 | 3 0.00161±0.00013 | 3 0.00128±0.00002 | 3 0.00162±0.00013 |
| 185 | 8 | 3 | 5 | Ti | Granite | ao | 3 0.21541±0.01714 | 3 0.04786±0.00722 | 3 0.04649±0.00726 | 0.05701±0.00452 | 3 0.05549±0.00374 |
| 186 | 8 | 3 | 6 | Ti | Granite | unextracted | 3 99.78013±0.01716 | 3 99.95062±0.00717 | 3 99.95189±0.00712 | 99.94142±0.00454 | 3 99.94262±0.00386 |
| 187 | 8 | 4 | 1 | Ti | Schist | shoot | | | | 3 0.00003±0.00001 | 3 0.00001±0.00000 |
| 188 | 8 | 4 | 2 | Ti | Schist | root | | | | 3 0.00022±0.00002 | 3 0.00018±0.00006 |
| 189 | 8 | 4 | 3 | Ti | Schist | water | | 3 0.00006±0.00000 | 3 0.00005±0.00000 | 3 0.00007±0.00001 | 3 0.00008±0.00001 |
| 190 | 8 | 4 | 4 | Ti | Schist | aa | 3 0.00346±0.00025 | 3 0.00039±0.00016 | 3 0.00038±0.00006 | 3 0.00116±0.00011 | 3 0.00093±0.00011 |
| 191 | 8 | 4 | 5 | Ti | Schist | ao | 3 0.07117±0.00206 | 3 0.00868±0.00151 | 3 0.01932±0.01151 | 0.04384±0.00097 | 3 0.02450±0.00806 |
| 192 | 8 | 4 | 6 | Ti | Schist | unextracted | 3 99.92537±0.00231 | 3 99.99100±0.00152 | 3 99.98025±0.01145 | 99.95469±0.00209 | 3 99.97426±0.00826 |
| 193 | 9 | 1 | 1 | Mn | Basalt | shoot | | | | 3 0.00174±0.00069 | 3 0.00306±0.00068 |
| 194 | 9 | 1 | 2 | Mn | Basalt | root | | | | 3 0.00148±0.00030 | 3 0.00232±0.00043 |
| 195 | 9 | 1 | 3 | Mn | Basalt | water | | 3 0.00081±0.00000 | 3 0.00057±0.00031 | 3 0.00057±0.00031 | 3 0.00222±0.00115 |
| 196 | 9 | 1 | 4 | Mn | Basalt | aa | 3 0.13394±0.00102 | 3 0.12103±0.00163 | 3 0.13458±0.00134 | 0.12768±0.00078 | 3 0.13265±0.01060 |
| 197 | 9 | 1 | 5 | Mn | Basalt | ao | 3 5.19660±0.05481 | 2.94864±0.13380 | 3 3.26902±0.11890 | 3.09991±0.01777 | 3 3.50626±0.21153 |
| 198 | 9 | 1 | 6 | Mn | Basalt | unextracted | 3 94.66946±0.05423 | 3 96.92992±0.13471 | 3 96.59559±0.11861 | 96.76862±0.01978 | 3 96.35349±0.21217 |
| 199 | 9 | 2 | 1 | Mn | Rhyolite | shoot | | | | 3 0.04278±0.00966 | 3 0.05212±0.00666 |
| 200 | 9 | 2 | 2 | Mn | Rhyolite | root | | | | 3 0.03312±0.00601 | 3 0.03983±0.00389 |
| 201 | 9 | 2 | 3 | Mn | Rhyolite | water | | 3 0.07167±0.05279 | 3 0.02557±0.00483 | 3 0.04063±0.01828 | 3 0.03651±0.00292 |
| 202 | 9 | 2 | 4 | Mn | Rhyolite | aa | 3 3.26617±0.42036 | 2.45942±0.55968 | 3 1.72494±0.09028 | 1.68700±0.20006 | 1.80434±0.17568 |
| 203 | 9 | 2 | 5 | Mn | Rhyolite | ao | 3 45.65611±1.39764 | 37.77452±1.60405 | 3 20.60621±2.53077 | 19.06043±2.95558 | 35.23366±2.77762 |
| 204 | 9 | 2 | 6 | Mn | Rhyolite | unextracted | 3 51.07772±1.04351 | 59.69439±2.20426 | 3 77.64328±2.61829 | 79.13604±3.15928 | 62.83354±2.67419 |
| 205 | 9 | 3 | 1 | Mn | Granite | shoot | | | | 3 0.00527±0.00097 | 3 0.00690±0.00109 |
| 206 | 9 | 3 | 2 | Mn | Granite | root | | | | 3 0.01037±0.00325 | 3 0.00768±0.00103 |
| 207 | 9 | 3 | 3 | Mn | Granite | water | | 3 0.01169±0.00123 | 3 0.00606±0.00271 | 3 0.01059±0.00418 | 3 0.01188±0.00165 |
| 208 | 9 | 3 | 4 | Mn | Granite | aa | 3 1.02515±0.04061 | 3 0.53762±0.08178 | 3 0.55199±0.08959 | 0.41244±0.04945 | 3 0.52635±0.09088 |
| 209 | 9 | 3 | 5 | Mn | Granite | ao | 3 1.29002±0.08306 | 0 | 3 0.21622±0.12600 | 1.09630 | 1.00938±0.08345 |
| 210 | 9 | 3 | 6 | Mn | Granite | unextracted | 3 97.68483±0.09942 | 3 99.45069±0.08200 | 3 99.22573±0.19068 | 99.46503±0.06950 | 3 98.43781±0.42806 |
| 211 | 9 | 4 | 1 | Mn | Schist | shoot | | | | 3 0.01884±0.00275 | 3 0.01973±0.00315 |
| 212 | 9 | 4 | 2 | Mn | Schist | root | | | | 3 0.01227±0.00187 | 3 0.01442±0.00283 |
| 213 | 9 | 4 | 3 | Mn | Schist | water | | 3 0.02384±0.00613 | 3 0.00328±0.00044 | 3 0.00141±0.00047 | 3 0.00217±0.00000 |
| 214 | 9 | 4 | 4 | Mn | Schist | aa | 3 0.11356±0.00262 | 3 0.08450±0.03069 | 3 0.02310±0.01361 | 3 0.01905±0.01576 | |
| 215 | 9 | 4 | 5 | Mn | Schist | ao | 3 0.34360±0.01230 | 3 0.02387 | | | |
| 216 | 9 | 4 | 6 | Mn | Schist | unextracted | 3 99.54284±0.01443 | 3 99.88370±0.04313 | 3 99.97362±0.01392 | 99.95518±0.01482 | 3 99.96368±0.00682 |
| 217 | 10 | 1 | 1 | Fe | Basalt | shoot | | | | 3 0.00008±0.00001 | 3 0.00010±0.00005 |
| 218 | 10 | 1 | 2 | Fe | Basalt | root | | | | 3 0.00015±0.00003 | 3 0.00023±0.00004 |
| 219 | 10 | 1 | 3 | Fe | Basalt | water | | 3 0.00001±0.00000 | 3 0.00000±0.00000 | 3 0.00001±0.00000 | 3 0.00001±0.00000 |
| 220 | 10 | 1 | 4 | Fe | Basalt | aa | 3 0.01490±0.00030 | 3 0.01437±0.00048 | 3 0.01649±0.00019 | 3 0.01625±0.00079 | 3 0.01630±0.00068 |
| 221 | 10 | 1 | 5 | Fe | Basalt | ao | 3 5.93598±0.03684 | 4.05949±0.14949 | 3 4.06576±0.45622 | 3.98648±0.09946 | 3 4.40022±0.41434 |
| 222 | 10 | 1 | 6 | Fe | Basalt | unextracted | 3 94.04912±0.03671 | 3 95.92614±0.14929 | 3 95.91775±0.45604 | 95.99709±0.10020 | 3 95.58313±0.41376 |
| 223 | 10 | 2 | 1 | Fe | Rhyolite | shoot | | | | 3 0.00013±0.00005 | 3 0.00013±0.00002 |
| 224 | 10 | 2 | 2 | Fe | Rhyolite | root | | | | 3 0.00026±0.00003 | 3 0.00038±0.00007 |
| 225 | 10 | 2 | 3 | Fe | Rhyolite | water | | 3 0.00004±0.00001 | 3 0.00004±0.00000 | 3 0.00007±0.00002 | 3 0.00002±0.00000 |

| | | | | | | | | | | | | | | | | |
|---|---|---|---|---|---|---|---|---|---|---|---|---|---|---|---|---|
| 226 | 10 | 2 | 4 | Fe | Rhyolite | aa | 3 | 0.03075±0.01792 | 3 | 0.03354±0.00151 | 3 | 0.04591±0.00335 | 3 | 0.03900±0.00351 | 3 | 0.03076±0.00200 |
| 227 | 10 | 2 | 5 | Fe | Rhyolite | ao | 3 | 1.95582±0.09841 | 3 | 1.97391±0.16405 | 3 | 1.86329±0.14344 | 3 | 1.78132±0.09922 | 3 | 1.90483±0.10493 |
| 228 | 10 | 2 | 6 | Fe | Rhyolite | unextracted | 3 | 98.01343±0.11459 | 3 | 97.99252±0.16325 | 3 | 98.09076±0.14037 | 3 | 98.17923±0.09618 | 3 | 98.06388±0.10328 |
| 229 | 10 | 3 | 1 | Fe | Granite | shoot | | | | | | | 3 | 0.00012±0.00001 | 3 | 0.00009±0.00003 |
| 230 | 10 | 3 | 2 | Fe | Granite | root | | | | | | | 3 | 0.00014±0.00004 | 3 | 0.00021±0.00004 |
| 231 | 10 | 3 | 3 | Fe | Granite | water | | | 3 | 0.00003±0.00000 | 3 | 0.00002±0.00000 | 3 | 0.00004±0.00001 | 3 | 0.00004±0.00001 |
| 232 | 10 | 3 | 4 | Fe | Granite | aa | 3 | 0.02562±0.00123 | 3 | 0.02143±0.00215 | 3 | 0.02210±0.00191 | 3 | 0.01753±0.00060 | 3 | 0.02100±0.00143 |
| 233 | 10 | 3 | 5 | Fe | Granite | ao | 3 | 5.47833±0.30748 | 3 | 4.30096±0.66925 | 3 | 4.01667±0.82253 | 3 | 5.24135±0.40284 | 3 | 4.94663±0.38957 |
| 234 | 10 | 3 | 6 | Fe | Granite | unextracted | 3 | 94.49606±0.30841 | 3 | 95.67758±0.67132 | 3 | 95.96120±0.82413 | 3 | 94.74083±0.40235 | 3 | 95.03203±0.39063 |
| 235 | 10 | 4 | 1 | Fe | Schist | shoot | | | | | | | 3 | 0.00006±0.00001 | 3 | 0.00007±0.00002 |
| 236 | 10 | 4 | 2 | Fe | Schist | root | | | | | | | 3 | 0.00027±0.00003 | 3 | 0.00027±0.00003 |
| 237 | 10 | 4 | 3 | Fe | Schist | water | | | 3 | 0.00010±0.00000 | 3 | 0.00007±0.00000 | 3 | 0.00007±0.00000 | 3 | 0.00010±0.00001 |
| 238 | 10 | 4 | 4 | Fe | Schist | aa | 3 | 0.01429±0.00060 | 3 | 0.00200±0.00028 | 3 | 0.00187±0.00026 | 3 | 0.00413±0.00057 | 3 | 0.00284±0.00048 |
| 239 | 10 | 4 | 5 | Fe | Schist | ao | 3 | 0.15184±0.00667 | 3 | 0.06373±0.01918 | 3 | 0.15232±0.06306 | 3 | 0.15681±0.01660 | 3 | 0.10039±0.02560 |
| 240 | 10 | 4 | 6 | Fe | Schist | unextracted | 3 | 99.83386±0.00712 | 3 | 99.93417±0.01915 | 3 | 99.84574±0.06301 | 3 | 99.83866±0.01713 | 3 | 99.89634±0.02611 |